\documentclass[fleqn,usenatbib]{mnras}

\usepackage{newtxtext,newtxmath}
\usepackage[T1]{fontenc}
\usepackage{ae,aecompl}

\usepackage{graphicx}	
\usepackage{amsmath}	
\usepackage{amssymb}	

\newcommand{\mr}{\mathrm}
\renewcommand{\vec}[1]{\boldsymbol{\mathbf{#1}}}
\usepackage{mathrsfs}

\usepackage{bm}
\usepackage{url}
\graphicspath{{figs/}}
\newcommand{\comm}[1]{}
\usepackage{enumerate}
\usepackage{pifont}
\usepackage{multirow}
\usepackage{color, colortbl}
\definecolor{WHITE}{RGB}{255, 255, 255}
\definecolor{HEADER}{RGB}{202,225,255}
\definecolor{I}{RGB}{180,251,180}
\definecolor{II}{RGB}{250,250,210}
\definecolor{III}{RGB}{255,229,204}
\definecolor{IV}{RGB}{255,228,225}

\usepackage{stackengine}

\usepackage{tabulary}
\newcolumntype{K}[1]{>{\centering\arraybackslash}p{#1}}

\usepackage{cellspace}
\usepackage{setspace}
\AtBeginEnvironment{tabular}{\singlespacing}

\title[{\sc Aura}: Retrieval of stellar and planetary properties]{Retrieval of planetary and stellar properties in transmission spectroscopy with {\sc Aura}}

\author[Pinhas et al.]{
Arazi Pinhas,$^{1}$\thanks{E-mail: \textcolor{blue}{ap817@ast.cam.ac.uk}} 
Benjamin V. Rackham,$^{2}\thanks{Earths in Other Solar Systems Team, NASA Nexus for Exoplanet System Science.}$
Nikku Madhusudhan,$^{1}$
D\'aniel Apai$^{2}$$^{,3}$\footnotemark[2]
\\
$^{1}$Institute of Astronomy, University of Cambridge, Madingley Road, CB3 0HA\\
$^{2}$Department of Astronomy/Steward Observatory, The University of Arizona, 933 N. Cherry Avenue, Tucson, AZ 85721, USA\\
$^{3}$Lunar and Planetary Laboratory, The University of Arizona, 1629 E University Boulevard, Tucson, AZ 85721, USA
}

\date{Accepted 2018 August 10. Received 2018 August 09; in original form 2018 April 04}
\pubyear{2018}

\begin{document}
\label{firstpage}
\pagerange{\pageref{firstpage}--\pageref{lastpage}}
\maketitle

\begin{abstract} 
Transmission spectroscopy provides a powerful probe of the atmospheric properties of transiting exoplanets. To date, studies of exoplanets in transit have focused on inferring their atmospheric properties such as chemical compositions, cloud/haze properties, and temperature structures. However, surface inhomogeneities in the host stars of exoplanets in the form of cool spots and hot faculae can in principle imprint signatures on the observed planetary transit spectrum. Here we present \textsc{Aura}, a new retrieval paradigm for inferring both planetary and stellar properties from a transmission spectrum. We apply our retrieval framework to a sample of hot giant exoplanets to determine the significance of stellar heterogeneity and clouds/hazes in their spectra. The retrieval analyses distinguish four groups of planets. First, the spectra of WASP-6b and WASP-39b are best characterised by imprints of stellar heterogeneity and hazes and/or clouds. HD 209458b and HAT-P-12b comprise the second group for which there is weak evidence for stellar heterogeneity and a high significance of hazes and/or clouds. The third group constitutes HAT-P-1b and WASP-31b and shows weak evidence against stellar heterogeneity but weak to substantial indications of clouds/hazes. The fourth group -- WASP-19b, WASP-17b, and WASP-12b -- is fit best by molecular and alkali absorbers with H$_2$ scattering without evidence for stellar heterogeneity and weak to no evidence for clouds/hazes. Our retrieval methodology paves the way to simultaneous information on the star and planet from higher resolution spectra using future facilities such as the James Webb Space Telescope and large ground-based facilities.  
\end{abstract}


\section{Introduction}\label{intro}

Transmission spectroscopy has been one of the most successful ways towards characterising exoplanetary atmospheres \citep[][]{burrows14,madhu16}. Studies of exoplanets in transit have been used to infer a wide variety of atmospheric properties including chemical compositions and abundances, clouds and hazes, and temperature profiles \citep[e.g.][]{sing16, kreidberg14a, kreidberg14b, knutson14a,knutson14b, madhu14, pont13, chen18, demory13, macdonald17, sedaghati18}. An essential assumption of most such studies is a homogeneous stellar photosphere characterized by one disk-integrated spectrum, and hence one stellar temperature and radius. However, stellar photospheres are not perfectly homogeneous. The stellar disk is generally comprised of differential areas each with a unique spectrum that can differ substantially from one representative average spectrum \citep{chapman87, shapiro14}.

Active regions of the stellar surface, in the form of cool spots or hot faculae, are among the primary features of stars responsible for heterogeneity of the photosphere and variability of the stellar brightness in time. Such features on active stars may factor significantly and can induce modifications to an otherwise pristine planetary transmission spectrum and hence retrieved atmospheric properties \citep{rackham17, oshagh14, mccullough14, pont13,barstow15}. The slope of a transmission spectrum in the visible wavelength region is usually interpreted as hazes and/or clouds composed of small particles, and yet can also be caused by cool star spots which are unocculted by the transiting planet. For example, the steep optical slope of HD 189733b's transmission spectrum has been interpreted as opacity from haze particles in the planetary atmosphere \citep{pont13} as well as a possible signature of activity features in the photosphere of its variable host star, either unocculted star spots \citep{mccullough14} or occulted stellar plages \citep[i.e. bright chromospheric regions,][]{oshagh14}. On the other hand, hot faculae or plages, when not occulted by the transiting planet, decrease the observed optical transit spectrum due to an increasing stellar radius at shorter wavelengths. The optical transmission spectrum of GJ 1214b, which displays a significant decline towards shorter wavelengths and shallower transit depths than observed in the near-infrared \citep{kreidberg14a}, has recently suggested a heterogeneous stellar photosphere dominated by hot faculae \citep{rackham17}. Unocculted faculae have also been suggested to affect the transmission spectrum of GJ 1132b, which displays a significant decrease in transit depth at optical wavelengths \citep{dittmann17} and, like GJ 1214b, orbits a mid-M dwarf star \citep{berta-thompson15}. The suite of effects issued by active stars relates the importance for a framework capable of simultaneously dissecting properties of the planetary atmosphere and heterogeneous stellar photosphere from an observed spectrum.

Here we present {\sc Aura}, a new retrieval framework that enables conjoint inference of exoplanetary and stellar properties. In addition to the usual properties explored for exoplanetary atmospheres (chemical abundances, clouds and hazes, and temperature structures), we incorporate a model that generally accounts for activity properties of the stellar photosphere \citep{rackham17}. The latter enables inferences on the fractional disk coverage of unocculted heterogeneity features, the average temperature thereof, and the temperature of the stellar photosphere. Our retrieval methodology ushers in the first instance of simultaneous retrieval of stellar and planetary properties constrained with present spectra, a consideration that will be integral to interpretations of future high fidelity observations from the James Webb Space Telescope (JWST) and large ground-based facilities.

Our work is presented as follows. We detail the components of our retrieval framework in Section \ref{retrieval_model} and demonstrate the self-consistency of the retrieval methodology in Section \ref{retrieval_consistency}. We apply the new methodology to a sample of hot giant exoplanets in Section \ref{retrieval_application} to determine the significance of stellar heterogeneity and clouds/hazes in their observed transmission spectra. Limitations of our approach and necessary future considerations are presented in Section \ref{sec:limitations}. We discuss future observing prospects and summarise in Section \ref{discussion_conclusions}.

\section{Retrieval Methodology}\label{retrieval_model}

\begin{figure}
\centering
\includegraphics[width = 0.47\textwidth]{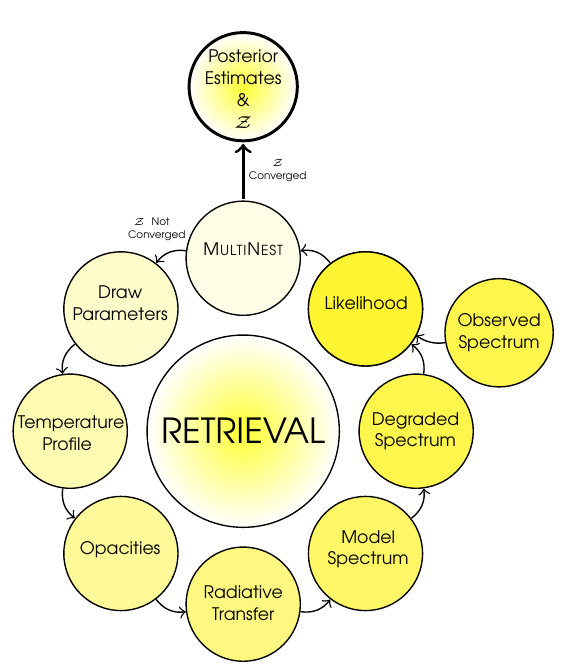}%
\caption{Illustrated retrieval methodology. Model parameters are drawn using the {\sc MultiNest} sampling algorithm and used to generate the temperature profile throughout the atmosphere. The temperature profile informs the relevant opacities which are used to evaluate the transfer of stellar radiation through the planetary atmosphere, producing a transmission spectrum that incorporates contamination from the heterogeneous stellar surface. The model transmission spectrum is convolved and binned, degrading the model spectrum to the quality of the observed spectrum. The relevance of the degraded model is then straightforwardly compared with the observed spectrum via the likelihood. The likelihoods are ascertained with the {\sc MultiNest} nested sampling algorithm which maneuvers through the space of model parameters and finds those with highest likelihoods. Once convergence is achieved on the `evidence' statistic ($\mathcal{Z}$), {\sc MultiNest} stops its iterative search and returns the final evidence and parameter estimates.}
\label{fig:retrieval_loop}
\end{figure}

We present a new Bayesian retrieval method, {\sc Aura}, to simultaneously infer properties of the planetary atmosphere and stellar photosphere from observed transmission spectra. Retrieving model parameters for the star and planet from transmission observations requires three basic components: a combined forward model of the planetary atmosphere and heterogeneous stellar photosphere, a statistical sampling method, and spectra which are uncorrected for stellar activity effects. We here outline the first two components in turn and describe the uncorrected data in Section \ref{uncorrected_data}. A condensed representation of the retrieval methodology is shown in Figure \ref{fig:retrieval_loop}.

\subsection{Forward Model}

\subsubsection{Pressure-Temperature Profile}\label{pT_profile}

The temperature as a function of height or pressure in the planetary atmosphere is calculated through the parametric relations in \citet{madhu09}. The parametric pressure-temperature ($p$--$T$) profile is motivated by physical principles and is able to fit diverse planetary atmosphere structures, mimicking atmospheric conditions of solar system planets as well as exoplanetary atmosphere models in the literature. The atmosphere is layered into three components with the following equations for the temperature, 

\begin{align}
    T &= T_0 + \left ( \frac{\mr{ln}(P/P_0)}{\alpha_1} \right )^2 \hspace{0.5cm} P_0 < P < P_1 \\
    T &= T_2 + \left ( \frac{\mr{ln}(P/P_2)}{\alpha_2} \right )^2 \hspace{0.5cm} P_1 < P < P_3 \\
    T &= T_2 + \left ( \frac{\mr{ln}(P_3/P_2)}{\alpha_2} \right )^2 \hspace{0.5cm} P > P_3  
\end{align}
A schematic of the parameterised profile is illustrated in Figure 1 of \citet{madhu09}. The free parameters of the temperature profile are $T_0$, the temperature at the top of the atmosphere; $\alpha_1$ and $\alpha_2$, values responsible for the gradient of the profile; and $P_1$, $P_2$, and $P_3$ which define the distinct atmosphere layers and generally determine the presence of a thermal inversion. The pressure at the top of the atmosphere is $P_0$. In the primary transit case the planetary source function is negligible; hence, there is no potential thermal inversion and the pressure parameters satisfy $P_2 \leq \, P_1 < P_3$. We partition our model atmosphere into 100 layers spaced equally in log-pressure between $10^{-6}$ bar and $10^{2}$ bar. While the lower level in the atmosphere of 100 bar is well below the observable exoplanet photosphere, the upper level of $10^{-6}$ bar marks the region where molecules are photo-dissociated and no longer contribute opacity \citep{moses11,moses13,moses14}. In addition, a seventh free parameter is the reference pressure $P_{\mr{ref}}$, the unknown prior pressure at $R_p$. The atmospheric temperature profile provides for the total number density $n_{\mr{tot}}(T)$ under the assumption that the gas is ideal.

\subsubsection{Sources of Chemical Opacity}

In addition to the temperature structure of the atmosphere, the retrieval forward model contains a suite of chemical species in the atmosphere that absorb or scatter incident light from the star. We consider several species expected to be prevalent in the atmospheres of hot Jupiters and with significant opacity in the spectral range of observations between 0.3 $\mu$m and 5 $\mu$m \citep{madhu12,moses2013,venot2015}. The species include Na, K, H$_2$O, CO, CH$_4$, CO$_2$, HCN, and NH$_3$. The volumetric mixing ratio of each species, $X_{i} = n_i / n_{\mr{tot}}$, is assumed uniform in the atmosphere and each is a free parameter in the retrieval framework. The volumetric mixing ratio of diatomic hydrogen is calculated assuming a H$_2$-He dominated atmosphere with a solar composition He/H$_2$ of 0.18 \citep[derived from][]{asplund09} and requiring the summation of relative abundances be unity, $X_{\mr{H_2}} = (1 - \Sigma_{i , i \neq \mr{He}} X_i)/ (1+X_{\mr{He}}/X_{\mr{H_2}})$.

We consider line absorption from each of the above species and collision-induced absorption (CIA) from H$_2$-H$_2$ and H$_2$-He. The molecular cross-sections for CH$_4$, HCN, and NH$_3$ are computed from EXOMOL line data \citep{tennyson16} and those for H$_2$O, CO, and CO$_2$ are obtained from HITEMP \citep{rothman10}. The CIA line data are sourced from the HITRAN archive \citep{richard12}. These line data are applied with a Voigt function to incorporate both temperature (Doppler) and pressure broadening. The computed molecular cross-sections are binned to a resolution of 0.01 cm$^{-1}$ on a pre-defined temperature and pressure grid spanning 10$^{-5}$ to 10$^{2}$ bar and 300 -- 3500 K. The established grid is interpolated to extract the cross-section for a general temperature, pressure and wavelength. A complete description of molecular cross-section calculations from the latest available line-list data is presented in \citet{gandhi17_genesis}.

\begin{figure}
\centering
\includegraphics[width = \columnwidth, trim= 0.2cm 0.2cm 0 0,clip]{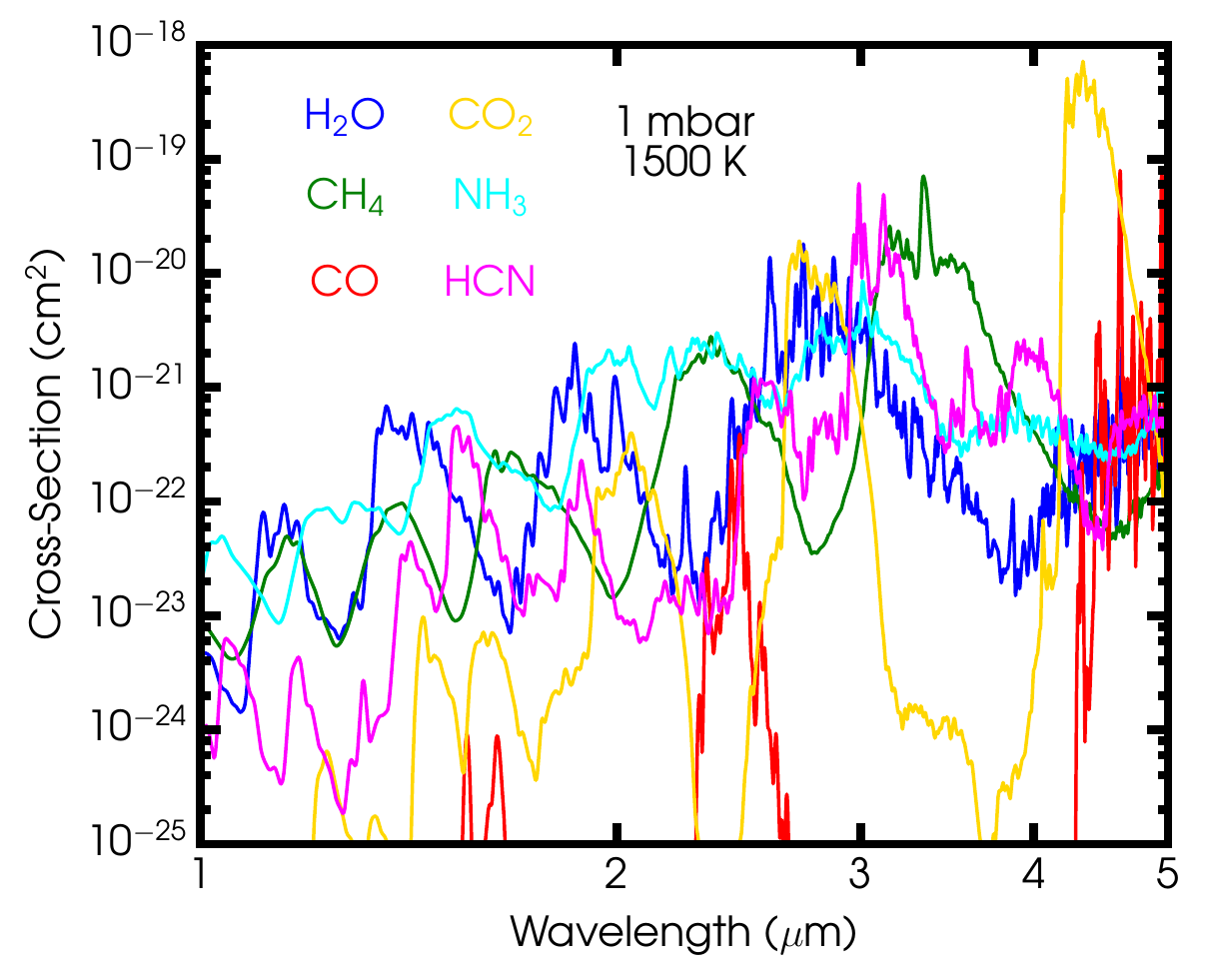}%
\caption{Cross-sections of molecular species considered in our forward model. These are shown at the approximate average equilibrium temperature of the hot giant exoplanets considered in this work and a pressure of 1 mbar, where the slant optical depth $\tau_{\mr{s}}$ is $\sim$1. The cross-sections are smoothed for clarity.}
\label{fig:opacs}
\end{figure}

The molecular cross-sections thus obtained are evaluated at each $p$--$T$ point of the atmosphere along the path of a ray. The contributions from all species are summed at each point and weighed by their mixing ratios $X_i$ to give the total extinction coefficient $\kappa$ at each point along a slant path $s$, 
\begin{align}
\kappa(p,T,\lambda) &= \sum_i X_i n_{\mr{tot}}(p,T) \, {\sigma_{i}(p,T,\lambda)} + X_{\mr{H_2}} n_{\mr{tot}}(p, T)\sigma_{\mr{H_2}} (\lambda) \label{extinc_coeff} \\
 + & \sum_{j}X_{\mr{H_2}}X_{j} n_{\mr{tot}}^2 \sigma_{\mr{H_2}-j}(T, \lambda), \nonumber \\
d\tau_s(p,T,\lambda) &= \kappa(p,T,\lambda) ds.\label{dtau}
\end{align}
Here, the first term of Equation~(\ref{extinc_coeff}) contributes extinction from molecular species and alkali metals, the second term represents H$_2$ Rayleigh scattering, and the third term serves for CIA extinction from interactions between H$_2$ and He such that $j \in \{\mr{H_2}, \mr{He}\}$. The CIA `cross-section' $\sigma_{\mr{H_2}-j}$ has units of [$\mr{m^{-1}amagat^{-2}}$] or [$\mr{m^5}$]. Figure \ref{fig:opacs} shows the computed molecular cross-sections for the species considered in this work at 1500 K and 1 mbar. The temperature represents the approximate average equilibrium temperature of the planets considered in this work.

\subsubsection{Cloud and Haze Model}

Observations of transiting exoplanets have suggested clouds and hazes across a wide spectrum of planetary types \citep[e.g.][]{kreidberg14a, knutson14a, sing16, macdonald17}. The terms `clouds' and `hazes' are typically used in specific contexts in the literature. From a formation standpoint, `hazes' imply particles formed through photochemical processes whereas a `cloud' constitutes particles formed through condensation of vapour on to a nucleus under suitable thermodynamic conditions \citep{marley13}. Hazes and clouds typically constitute small and large particles, respectively, where the distinction in particle size occurs at $\sim$0.1 $\mu$m.

These terms are also used in reference to the predominant morphology of spectral features they induce in transmission spectra, especially in parametric models used for atmospheric retrieval or forward models \citep[e.g.,][]{benneke12, kreidberg14b,sing16,macdonald17}. A `cloud' is generally used to mean a source of grey opacity from particle sizes $\gtrsim$1 $\mu$m throughout the spectral range effective to a certain height in the atmosphere, while a `haze' is represented by a non-grey opacity in the optical wavelength region and through a power-law dependence on wavelength \citep[e.g.][]{macdonald17}. In essence, hazes produce Rayleigh-like slopes in the visible spectral range while clouds produce grey opacity throughout the spectrum \citep{pinhas17}.

Our forward model includes a parameterization for inhomogeneous clouds and hazes which accounts for effects due to the broad range of particle sizes \citep{macdonald17}. The extinction coefficient (with units of inverse length) which broadly incorporates these two spectral effects is, 
\begin{equation}
    \kappa_{\mr{cloud/haze}} (r) = \left\{\begin{array}{ll}
    n_{\mr{H_2}} a \sigma_0(\lambda/\lambda_0)^{\gamma}\,\,\,\,\,\,\,\,\,\,{ \it if}\,\,P<P_{\mr{cloud}}\\
                  \infty\,\,\,\,\,\,\,\,\,\,\,\,\,\,\,\,\,\,\,\,\,\,\,\,\,\,\,\,\,\,\,\,\,\,\,\,\,\,\,\,\,\,\,\,\,\mr{otherwise}\\
                \end{array}
              \right.
\end{equation}
where the first relation represents a slope in the optical characteristic of hazes and the second equality represents an opaque grey opacity across all wavelengths characteristic of clouds of particle sizes $\gtrsim$1 $\mu$m. Here, $\lambda_0$ is a reference wavelength (0.35 $\mu$m), $\sigma_0$ is the gaseous H$_2$ Rayleigh scattering cross-section at $\lambda_0$ ($5.31\times 10^{-31}\mr{m^2}$), $a$ is the `Rayleigh-enhancement factor' and $\gamma$ is the `scattering slope'. The Rayleigh-enhancement factor effectively quantifies the offset level of the optical transmission spectrum. In principle, inherent variations in species' refractive indices can constrain different species through the value of $a$. Moreover, the optical slope can also be used to constrain species since each species has characteristic values of $\gamma$ \citep{pinhas17}. Three cloud/haze parameters for our retrieval are thus $a$, $\gamma$, and $P_{\mr{cloud}}$. The fourth cloud/haze parameter $\bar{\phi}$ describes the terminator-averaged cloud/haze contribution of a two-dimensional planetary atmosphere and enters into the planet's transit depth as
\begin{equation}\label{effective_transit_depth}
    \Delta_{\mr{planet}}(\lambda) = \bar{\phi} \Delta_{\mr{cloud/haze}} (\lambda)  + (1 - \bar{\phi}) \Delta_{\mr{clear}}(\lambda).
\end{equation}
where $\Delta_{\mr{cloud/haze}}$ ($\Delta_{\mr{clear}}$) are the transit depths computed with (without) incorporation of clouds/hazes. A terminator completely covered with clouds/hazes has $\bar{\phi} = 1$ while a clear atmosphere along the terminator has $\bar{\phi} = 0$. A terminator with $0<\bar{\phi}<1$ contains patchy or inhomogeneous clouds and hazes. 

\subsubsection{Radiative Transfer of Stellar Light}

As an exoplanet transits its host star, incident stellar light filters through the planetary atmosphere and experiences differential extinction across wavelength due to absorption and scattering. The observed transmission spectrum is modeled through consideration of pencil rays travelling through each layer in the atmosphere. The atmosphere is assumed to be in hydro-static equilibrium and composed of an ideal gas. The transit depth components in Equation (\ref{effective_transit_depth}) for the two-dimensional atmosphere model are computed through \citep[see][Appendix]{macdonald17},
\begin{equation}\label{eq:transit_depth}
    \Delta_{j}(\lambda) = \frac{R_{\mr p}^2 + 2 \int^{R_{\mr p} + H_{\mr A}}_{R_{\mr p}} b \, (1 - e^{-\tau_{s, j}(b, \lambda)}) \, db - 2 \int^{R_{\mr p}}_{0} b e^{-\tau_{s, j}(b, \lambda)} db}{R_{\star}^2}, 
\end{equation}
 Here, $j \in$ \{cloud/haze, clear\} and denotes the component in Equation~(\ref{effective_transit_depth}), $R_{\star}$ is the stellar radius, $H_{\mr{A}}$ is the maximal atmospheric altitude (i.e. at 1 $\mu$bar), $b$ is the impact parameter perpendicular to the line-of-sight, and $\tau_{s,j}$ is the total slant optical depth for component $j$ at $b$. As such, $\tau_{s,j}(b, \lambda)$ is the integral of Equation~(\ref{dtau}) over slant length $s$. The variable planetary radius with wavelength in Equation (\ref{eq:transit_depth}) can be understood physically. The first term is achromatic absorption from an opaque planetary radius. The second term accounts for chromatic extinction of incident light rays above the fiducial $R_{\mr{p}}$, and the third term accounts for rays whose atmospheric optical depth lies below the assumed fiducial radius $R_{\mr{p}}$.

\subsubsection{Heterogeneous Stellar Photosphere}\label{het_star}

The photospheres of stars display heterogeneity in the form of magnetic active regions. Beyond observations of the Sun, the most straightforward evidence for this comes from observed periodic brightness variations \citep[e.g.][]{mcquillan14,newton16}, which correspond to active regions rotating into and out of the projected stellar disk. These observations, however,  place only lower limits on active region covering fractions, as longitudinally symmetric components of the surface heterogeneity do not contribute to brightness variations \citep{rackham18}.

The types of stellar heterogeneity considered in this work are cool spots and hot faculae, respectively dimmer and brighter areas of a star that evolve on timescale of days to weeks \citep[see][and references therein]{hathaway15}. The model for heterogeneous stellar photospheres used in the present work was developed previously and used to interpret the optical transmission spectrum of GJ 1214b \citep{rackham17}. The model is used concomitantly with the usual exoplanetary atmosphere properties to study the stellar and planetary nature of transmission spectra.

We utilize the {\it Composite Photosphere and Atmospheric Transmission} (CPAT) model to formally account for inhomogeneities in the stellar photosphere \citep{rackham17}. The model divides the stellar surface into two components, called the `unocculted' and `occulted' components, defined so by the region of the star occulted by the planet during transit. The occulted component of the stellar disk has an average spectral energy distribution $\mathcal{S}_{o}(\lambda)$ and contains the planetary transit chord. The unocculted component accounts for heterogeneity features (cool spots and/or hot faculae) outside of the transit chord in aggregate and has an average spectral energy distribution $\mathcal{S}_{u}(\lambda)$. We use PHOENIX atmospheric models \citep{husser13} to construct the two spectral components of the star. The spectral energy distributions are interpolated from the grid of PHOENIX models through specification of three stellar parameters including the photospheric temperature, stellar metallicity, and stellar gravity. In this analysis, we fix the stellar metallicity and gravity to literature values and parameterize the different components by temperature. The family of PHOENIX stellar models are limited to effective temperatures of $\sim$2300 K and hence we have incorporated the DRIFT-PHOENIX grid of atmospheric models \citep{witte11} which allows consideration of lower temperatures to 1,000 K. While both model grids are based on the PHOENIX stellar atmosphere code \citep{hauschildt99}, the DRIFT-PHOENIX models include additional physics applicable to cooler atmospheres, including the growth and settling of dust grains \citep{woitke03,woitke04,helling08a,helling08b,helling06,witte09}. While the host stars considered in this work have effective temperatures above 4,500 K, DRIFT-PHOENIX can be used to study present and future transmission spectra of planets orbiting cooler active stars.

\begin{figure*}
\centering
\includegraphics[width=\textwidth]{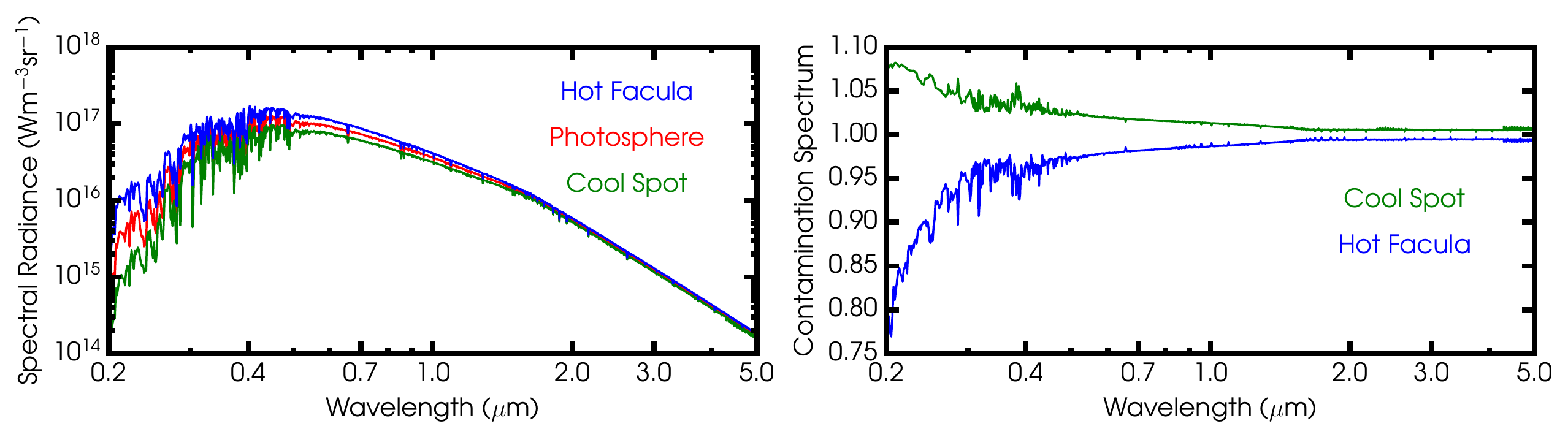}%
\caption{Spectral influence of stellar heterogeneity features. Left panel: Spectral radiances of a hot facula, the photosophere, and a cool spot. It is clear that $\mathcal{S}_{u}/\mathcal{S}_{o}$ increases (decreases) for a hot facula (cool spot) with bluer wavelengths. The spectrum of the photosphere was produced assuming properties of HAT-P-1b and adding (subtracting) 300 K for the hot facula (cool spot) spectrum. Right panel: Stellar contamination spectrum $\mathcal{E}_{\mr{het}}$. A cool spot (hot facula) displays an increase (decrease) in the contamination spectrum (and hence ultimately the observed spectrum) with decreased wavelengths. In both cases a covering fraction of ten percent is assumed.}
\label{fig:stel_components}
\end{figure*}

The observed planetary transmission spectrum accounting for heterogeneous stellar photospheres is
\begin{align}\label{observed_TD}
\hspace{1cm}
 \underbrace{\left ( \frac{\mathcal{R}_{p}(\lambda)}{\mathcal{R}_{\star}(\lambda)} \right)^2}_{\text{Observed}\atop\text{Spectrum}}
 & =  \underbrace{\left ( \frac{R_p(\lambda)}{R_{\star}} \right)^2}_{\text{Planetary}\atop\text{Spectrum}}
 \underbrace{\left ( 1-\delta \left ( 1 - \frac{\mathcal{S}_{u}(\lambda)}{\mathcal{S}_{o}(\lambda)}\right )\right )^{-1}}_{\text{Stellar}\atop\text{Contamination}} \\
 \text{or} & \mr{\,\,} \text{succinctly,} \nonumber \\
  \Delta_{\mr{obs}}
 & =  \Delta_{\mr{planet}} \mathcal{E}_{\mr{het}}.
 \end{align}
 Here, $\delta$ is the areal fraction of the projected stellar disk covered with heterogeneities (cool spots and/or hot faculae), $\mathcal{S}_{u}(\lambda)$ is the representative spectrum of the heterogeneities, and $\mathcal{S}_{o}(\lambda)$ is the spectrum of the occulted stellar surface, which we assume to be the immaculate photosphere. Equation (\ref{observed_TD}) can be understood plainly. The first term on the right is the `pristine' transit spectrum which alone considers effects from the planetary atmosphere, i.e. Equation (\ref{effective_transit_depth}). The second is a perturbative term that incorporates heterogeneity of the stellar disk and is the `contamination' spectrum $\mathcal{E}_{\mr{het}}$. The observed spectrum is the product of the two components. In the case of cool spots, $\mathcal{S}_{u}/\mathcal{S}_{o}$ decreases with bluer wavelengths, decreasing the apparent radius of the star $R_{\star}$ and hence increasing the observed transit depth $\Delta_{\mr{obs}}$. On the other hand, hot faculae increase $\mathcal{S}_{u}/\mathcal{S}_{o}$ with bluer wavelengths, thereby increasing the apparent radius of the star and decreasing $\Delta_{\mr{obs}}$. We note that both cool spots and hot faculae may be present simultaneously in the unocculted component of the stellar disk and thus jointly affect the transmission spectrum. However, these features produce opposing effects on the transmission spectrum, as illustrated in Figure~\ref{fig:stel_components}, and in practice, one effect will dominate the observed spectrum. In this light, $\delta$ broadly represents the relative overabundance of the dominant heterogeneity in the unocculted disk. Nevertheless, it is important to emphasize that Equation (\ref{observed_TD}) may be easily expanded to incorporate spot and facula effects separately as is expressed in Equation (3) of \citet{rackham18}.

\begin{figure*}
\centering
\includegraphics[scale=0.8]{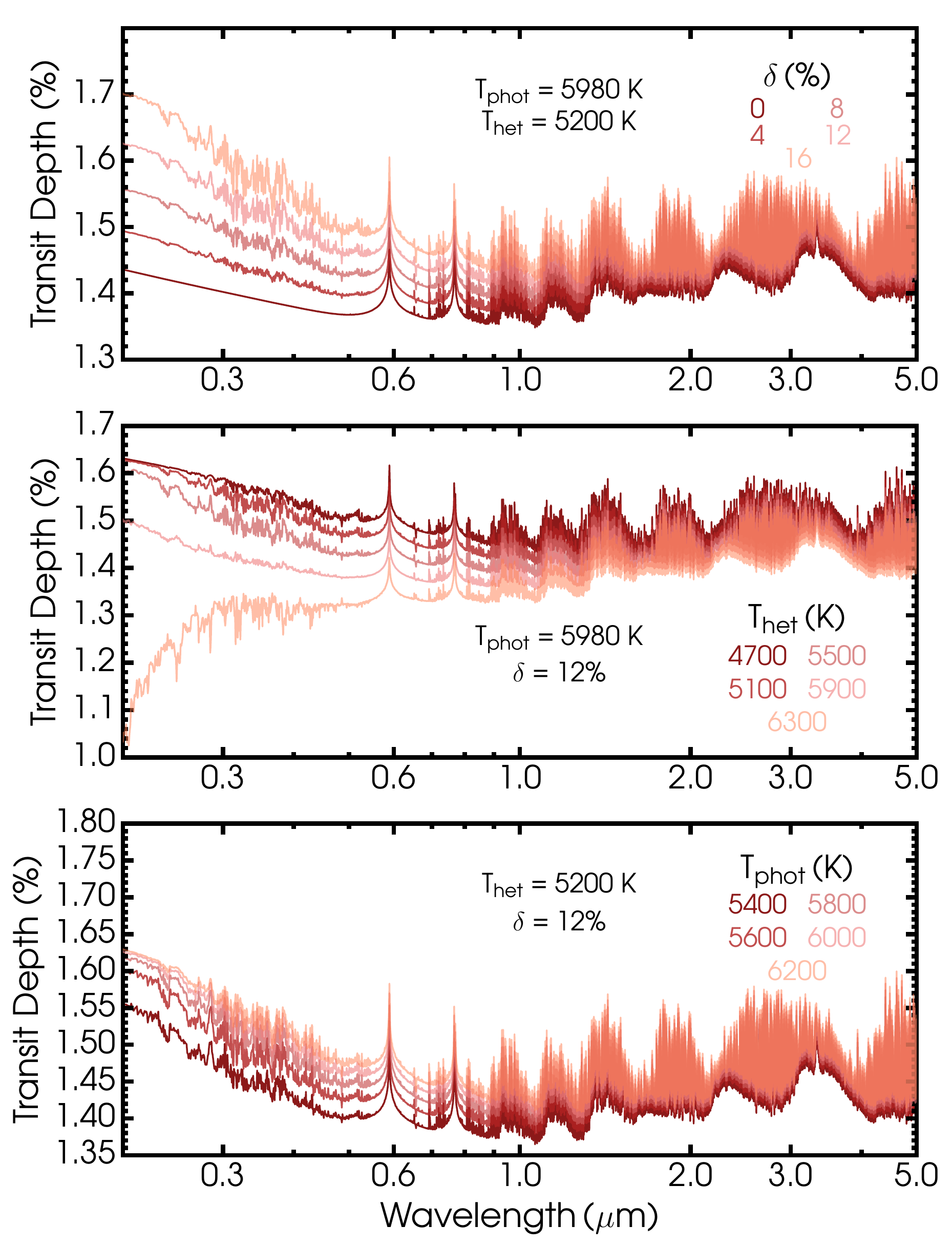}%
\caption{Effects of stellar properties on observed transmission spectra. The transmission spectra are calculated according to Equation~(\ref{observed_TD}) for underlying properties representative of HAT-P-1b. Each panel shows the influence of one stellar parameter. The upper, middle, and bottom panels illustrate variations in the covering fraction, heterogeneity temperature, and photospheric temperature, respectively. Lighter colours represent larger values in the considered parameters. The darker models do not show less variation in near-infrared molecular features; rather, these features are hidden by the lighter models.}
\label{fig:het_effects}
\end{figure*}

The variational effects of the three stellar parameters on observed transit depths are shown in Figure~\ref{fig:het_effects}. Sensitivity in all three parameters is greatest in the 0.2 $-$ 1.0 $\mu$m range. An increase in the heterogeneity covering fraction $\delta$ amplifies the ratio of spectral energy distributions, $\mathcal{S}_{u}/\mathcal{S}_{o}$, and therefore increasingly portrays the fine spectral details from the two stellar components. For a covering fraction dominated by cool spots, an increased $\delta$ amounts to a decreased apparent stellar radius and leads to higher values of the transit depth. The increase in transit depth is not uniform, with the visible spectral range affected much more than the near- and mid-infrared ranges. In addition, an increase in the heterogeneity temperature $T_{\mr{het}}$ modifies the fine spectral features apparent in the spectrum and is representative of different stellar types, as seen in the middle panel of Figure~\ref{fig:het_effects}. For $T_{\mr{het}}$ greater than $T_{\mr{phot}}$, the optical spectrum curves downward due to an increasing $\mathcal{S}_{u}/\mathcal{S}_{o}$ with bluer wavelengths, increasing the apparent radius of the star. As shown in the bottom panel, increasing the representative photospheric temperature $T_{\mr{phot}}$ dampens spectral features peculiar to the unocculted component and results in higher transit depths throughout the whole spectral range. 
 
In its entirety, our retrieval forward model contains 22 free parameters: seven for the pressure-temperature structure ($T_0$, $\alpha_1$, $\alpha_2$, $P_1$, $P_2$, $P_3$, $P_{\mr{ref}}$), eight for molecular and elemental abundances ($X_{\mr{Na}}$, $X_{\mr{K}}$, $X_{\mr{H_2O}}$, $X_{\mr{CH_4}}$, $X_{\mr{NH_3}}$, $X_{\mr{HCN}}$, $X_{\mr{CO}}$, $X_{\mr{CO_2}}$), four for clouds and hazes ($a$, $\gamma$, $P_{\mr{cloud}}$, $\bar{\phi}$), and three for the star ($\delta$, $T_{\mr{het}}$, $T_{\mr{phot}}$). 

\subsubsection{Model-Data Comparison}\label{model_data_comparison}

With a forward model established from components \ref{pT_profile} - \ref{het_star} for a set of parameter values, it is necessary to ascertain its degree of fit to observations. The statistical comparison of a generated high-resolution forward model to observations is a two-fold process. The high-resolution spectrum generated from (at most) 22 free parameters is first convolved with the point-spread function (PSF) of each instrument, which degrades the model abstraction to the instrument's capability. The convolved model spectrum is then further degraded through its binning to the spectral resolution of the data. 

In practice, the transition of a high-resolution model $\mathcal{M}$ to a lower-resolution convolved spectrum is calculated from
\begin{equation}
\mathcal{M}_{\mr{conv}}(\lambda) = \int^{\lambda_{\mr{max}}}_{\lambda_{\mr{min}}} \mathcal{M} (\lambda ') \mr{PSF}(\lambda - \lambda') d\lambda' .
\end{equation}
The convolved spectrum is then discretized to the resolution of the data through weighting by the instrument sensitivity function $s(\lambda)$, 
\begin{equation}
    \mathcal{M}_{\mr{binned}, i} = \frac{\int^{\lambda_{\mr{max},i}}_{\lambda_{\mr{min},i}} \mathcal{M}_{\mr{conv}} s(\lambda) d\lambda}{\int^{\lambda_{\mr{max},i}}_{\lambda_{\mr{min},i}} s(\lambda) d\lambda},
\end{equation}
This calculation is performed for each datum $i$ where $\lambda_{\mr{min},i}\,(\lambda_{\mr{max},i})$ is the minimum (maximum) wavelength value of the same.

\subsection{Bayesian Inference}

The essence of our retrieval approach is to use spectra of transiting exoplanets to infer various properties of the exoplanetary atmosphere and heterogeneous properties of the star. In addition to parameter estimation of planetary and stellar properties, it is necessary to compare the adequacy of different model realizations to the observations. Hence, we here adopt a Bayesian statistics approach towards estimation of planetary and stellar properties and model comparison.

The Bayesian algorithm used in this work is the {\sc MultiNest} multi-modal nested sampling technique that enables efficient calculation of the Bayesian evidence as well as parameter estimation \citep{skilling04, skilling06, feroz08, feroz09, feroz13}. The sampling procedure takes transit spectra as input and samples the complex multi-dimensional space of forward model parameters (planetary and stellar) to compute the evidence $\mathcal{Z}$. In addition, the nested sampling approach allows for joint and marginalised posterior distributions of the forward model parameters and their statistical credibility intervals. We implement the {\sc MultiNest} nested sampling algorithm through a Python wrapper, PyMultiNest \citep{buchner14}. The reader is referred to \citet{skilling04}, \citet{skilling06}, \citet{feroz13}, and \citet{buchner14} for extensive discussions on the nested sampling method. 

\subsubsection{Parameter Estimation}

The heart of Bayesian parameter estimation lies in one relation known as Bayes' Theorem. Consider a forward model $\mathcal{M}_i$ defined by a set of free parameters \vec{$\theta$}. The {\it a priori} values and probabilities of the parameters are described through the prior probability density function, $\upi(\vec{\theta} | \mathcal{M}_i)$. Consideration of observations allows for a formal modification on the probabilities of prior parameters through Bayes' Theorem written as, 
\begin{align}\label{bayes'_theorem}
p(\vec{\theta} | \mathcal{D}_{\mr{obs}}) & = \frac{\mathcal{L}(\mathcal{D}_{\mr{obs}} | \vec{\theta}, \mathcal{M}_i) \, \upi(\vec{\theta} | \mathcal{M}_i)}{\mathcal{Z}(\mathcal{D}_{\mr{obs}} | \mathcal{M}_i)}.
\end{align}
Here, the initial parameter plausabilities ($\upi(\vec{\theta} | \mathcal{M}_i)$) transform according to the plausability of the observations in light of the parameter values $\vec{\theta}$ ($\mathcal{L}(\mathcal{D}_{\mr{obs}} | \vec{\theta}, \mathcal{M}_i)$) to give a final distribution of probabilities for each parameter ($p(\vec{\theta} | \mathcal{D}_{\mr{obs}})$). Normalization of the posterior probability distribution is ensured through $\mathcal{Z}(\mathcal{D}_{\mr{obs}} | \mathcal{M}_i)$. A measure of fit between observations and forward model parameter values is given by the likelihood function as 
\begin{equation}
    \mathcal{L}(\mathcal{D}_{\mr{obs}} | \vec{\theta}, \mathcal{M}_i) = \prod_{k}^{N_{\mr{obs}}} (2\upi\sigma_k^2)^{-1/2} e^{-\frac{(\mathcal{D}_{\mr{obs},k} - \mathcal{M}_{i, k})^2}{2\sigma_k^2}},
\end{equation}
in which $\mathcal{M}_{i,k}$ is the $k^{th}$ binned datum of spectrum $\mathcal{M}_i$, as outlined in Section \ref{model_data_comparison}, and $\sigma_k$ is the precision on the $k^{th}$ datum. A determination of parameter values through Equation (\ref{bayes'_theorem}) does not in itself require a calculation of $\mathcal{Z}$ since the latter is merely a constant of normalization. However, this assumes the model -- among other putative models -- is best suited to describe the data. Therefore, a determination of $\mathcal{Z}$ is powerful when one aims to compare models with different assumptions (e.g., stellar heterogeneity versus no stellar heterogeneity). The evidence $\mathcal{Z}$, so called for its utility in providing evidence for one model over another, is a weighted sum of the likelihood function over the prior space, 
\begin{equation}
    \mathcal{Z} = \int_{\vec{\theta}} \mathcal{L}(\mathcal{D}_{\mr{obs}} | \vec{\theta}, \mathcal{M}_i) \, \upi(\vec{\theta} | \mathcal{M}_i) d \vec{\theta}.\label{evidence}
\end{equation}

\subsubsection{Model Comparison}

In addition to the parameter estimation achieved through Bayes' Theorem, the nested sampling method allows for a direct computation of $\mathcal{Z}$ and enables model comparisons. This is in contrast to Markov chain Monte Carlo methods which have been designed to obtain posterior distributions without providing $\mathcal{Z}$. The key to an efficient calculation of $\mathcal{Z}$ lies in collapsing the integral in Equation (\ref{evidence}) to a one-dimensional sum \citep{skilling06}. To obtain $\mathcal{Z}$ in this form, first we consider the cumulant prior mass defined by
\begin{equation}
    X(\beta) = \int_{\mathcal{L}(\vec{\theta})>\beta} \upi (\vec{\theta}) d\vec{\theta}
\end{equation}
and which considers all likelihood values greater than $\beta$. As $\beta$ increases the enclosed cumulant mass $X$ decreases from 1 to 0, where for $\beta = L_{\mr{max}}$, $X = 0$. Hence the evidence in Equation (\ref{evidence}) transforms to an integral across one dimension \citep{skilling06},
\begin{equation}
    \mathcal{Z} = \int_0^1 \mathcal{L}(X) dX,
\end{equation}
which is computed numerically through a weighted sum over all sets of sampled priors,
\begin{equation}
    \mathcal{Z} = \sum_i^{N_{\mr{sampled}}} \mathcal{L}_i w_i. \label{evidence_numerical}
\end{equation}

The nested sampling procedure runs with an evolving set of $N$ `live points' drawn from the prior mass where $N$ can be large for accuracy or small for expedition \citep{skilling06}. At each sampling iteration one live point with the lowest likelihood value $\mathcal{L}_{\mr{min}}$ is substituted with another point of likelihood $\mathcal{L}_{\mr{sub}}$ drawn from the prior distribution such that $\mathcal{L}_{\mr{sub}}>\mathcal{L}_{\rm min}$. By induction, the live points are drawn from steadily contracting iso-likelihood contours with progressive iterations. The sampling process continues as above with $\mathcal{Z}$  calculated by Equation (\ref{evidence_numerical}) and is terminated once a pre-figured tolerance on $\Delta \mathcal{Z}$ is achieved, providing for a final value and error on $\mathcal{Z}$.

With the evidence thus acquired, a quantitative comparison of two models is given through the ratio of their evidences \citep{trotta08}, 
\begin{equation}
    \mathcal{B}_{ij} = \frac{\mathcal{Z}_i}{\mathcal{Z}_j}
\end{equation}
where the sub-scripts $i$ and $j$ represent model $i$ and $j$, respectively. A $\mathcal{B}_{ij} > 1$ represents preference for $\mathcal{M}_i$ over $\mathcal{M}_j$. In particular, $\mathcal{B}_{ij}$ values of $1 - 3$, $3 - 20$, $20 - 150$, and >$150$ can be interpreted as `weak', `substantial', `strong', and `very strong' preferences of model $\mathcal{M}_i$ against model $\mathcal{M}_j$, respectively \citep{kass95}. This enables direct model comparison for distinct model realizations (e.g., stellar heterogeneity versus no stellar heterogeneity). In this work we assume uniform priors on the tested models, such that no one model is inherently preferred over another (however, see Sections 5.1 and 5.2). As discussed in \citet{trotta08} and \citet{benneke13}, the Bayes factor can be transformed into a detection significance (i.e. $\sigma$ significance) in the language of frequency statistics and effectively quantifies the significance of model $\mathcal{M}_i$ against $\mathcal{M}_j$. We communicate both statistics in the presentation of our results to accommodate readers familiar with either area.

\section{Retrieval Self-Consistency}\label{retrieval_consistency}

\begin{table}
\centering
\caption{Prior information used in retrieval analyses.} 
\label{tab:ret_priors}
\begin{tabular}{K{1.4cm}  K{1.8cm} K{3.2cm}}
\hline
\hline
\rule{0pt}{3ex}  
Parameter & Prior Distribution & Prior Range\\ [1ex]
\hline
\rule{0pt}{3ex}
$T_0$ & Uniform & 800 $-$ $T_{\mr{eq}}+200$ K (see Table \ref{tab:system_properties} for $T_{\mr{eq}}$) \\
$\alpha_{1,2}$ & Uniform & $0.02 - 1$ K$^{-1/2}$ \\
$P_{1,2}$ & Log-uniform & $10^{-6} - 10^{2}$ bar\\
$P_{3}$ & Log-uniform & $10^{-2} - 10^{2}$ bar\\
$X_i$ & Log-uniform & $10^{-12} - 10^{-2}$ \\
$a$ & Log-Uniform & $10^{-4} - 10^{8}$ \\
$\gamma$ & Uniform & -$20 - 2$ \\
$P_{\mr{cloud}}$ & Log-Uniform & $10^{-6} - 10^{2}$ bar\\
$\bar{\phi}$ & Uniform & $0 - 1$ \\
$\delta$ & Uniform & $0 - 0.5$ \\
$T_{\mr{het}}$ & Uniform & 0.5$T_{\mr{phot}} - 1.2T_{\mr{phot}}$\\
$T_{\mr{phot}}$ & Gaussian & Planet specific (see Table \ref{tab:system_properties})\\
\hline
\end{tabular}
\end{table}

\begin{figure*}
\centering
\includegraphics[scale=0.93]{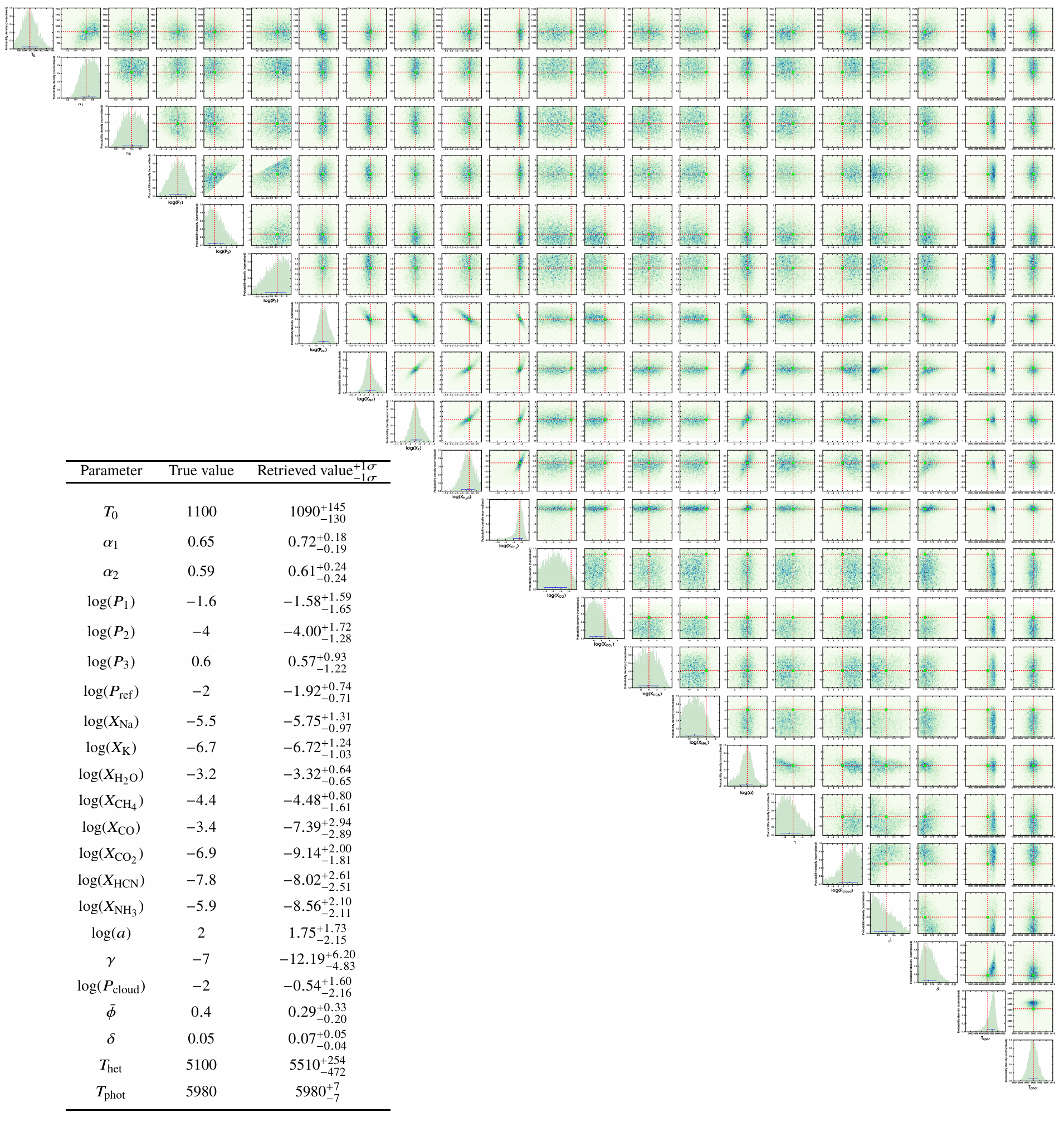}%
\caption{Posterior distributions of planetary and stellar properties for simulated observations. Shown in the main are joint and marginalised retrieved distributions. The red lines and green squares indicate the true values of the synthetic model used to generate the simulated data. The fully marginalized panels along the diagonal show retrieved median and 1$\sigma$ confidence intervals in blue. The inset table lists the true parameter values of the synthetic model and the retrieved parameter values with the 1$\sigma$ confidence limits.}
\label{fig:simret}
\end{figure*}

\begin{figure*}
\centering
\includegraphics[scale = 0.68]{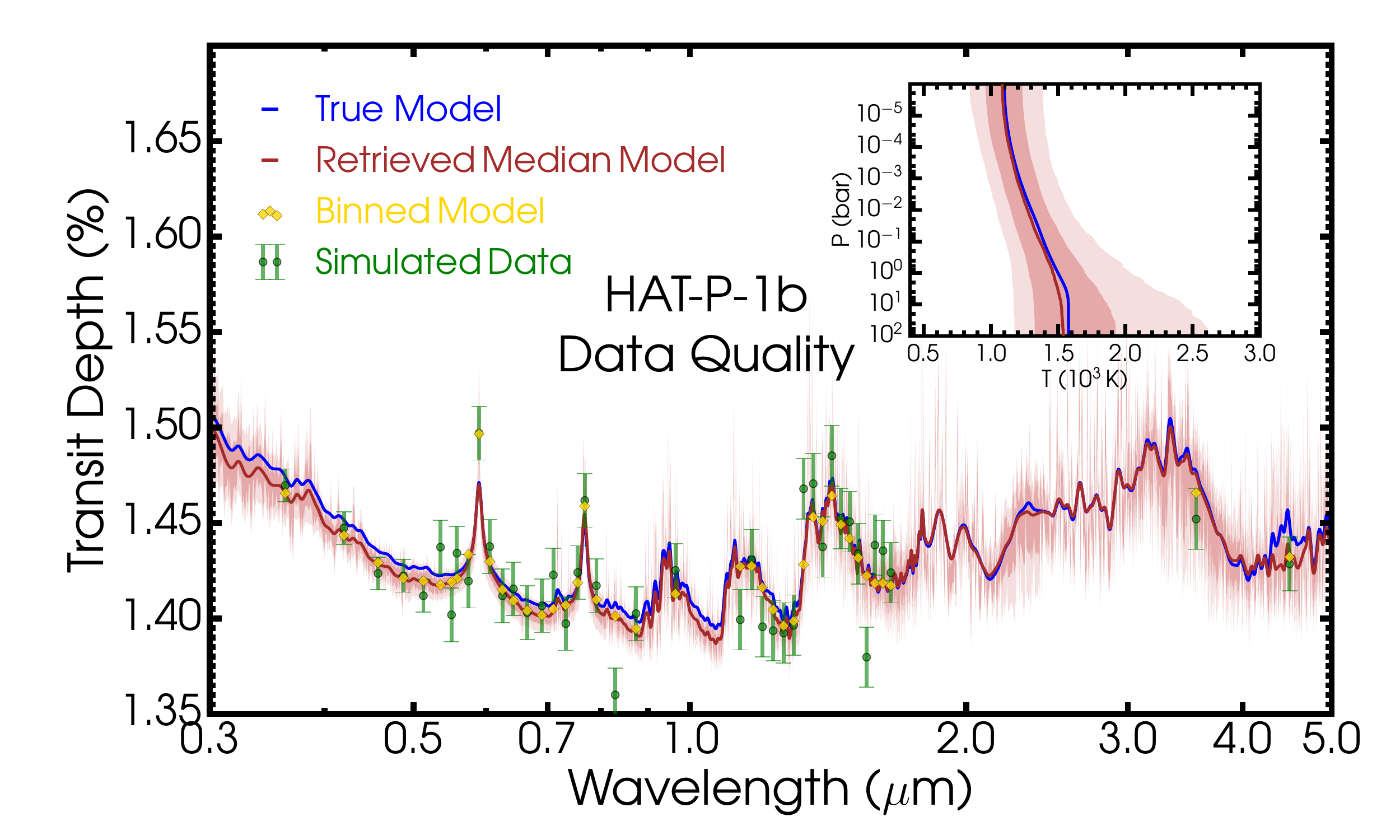}%
\caption{Retrieved spectrum and $p-T$ profile of the synthetic observations. Main Figure: The retrieved median spectrum and true spectrum used to generate the synthetic observations are shown in red and blue, respectively. Both models are smoothed for clarity. The light red envelopes around the median spectrum show the associated 1$\sigma$ and 2$\sigma$ confidence intervals. The simulated observations used as input to the retrieval are shown in green. The yellow points are the binned model points of the median spectrum at the same resolution as the observations. Inset Figure: The retrieved temperature profile of the simulated observations. As in the main figure, the red and blue spectra are the median retrieved and true $p-T$ profiles, respectively. The light red contours are the 1$\sigma$ and 2$\sigma$ confidence intervals of the retrieved profiles.}
\label{fig:fit_pT}
\end{figure*}

We present a retrieval framework for transmission spectra that enables joint estimation of planetary and stellar properties. Before an application to the spectra of a sample of hot Jupiters, we here examine the fidelity of our framework on simulated observations.

We use HAT-P-1b as a representative test case. The set of data for HAT-P-1b is of moderate quality among the studied hot Jupiters and is characterised by median spectral resolution and precisions. HAT-P-1 also shows an intermediate amount of stellar activity based on emission from singly-ionised calcium H and K lines \citep[i.e. log($R^{'}_{\mr{HK}}$) index,][]{noyes84, hall08} which is thought to be a proxy for stellar activity given its direct relationship with active regions in the Sun \citep{baliunas95}. The simulated exercise therefore examines the degree to which heterogeneity properties of active stars and atmospheric properties of exoplanets can be inferred in light of moderate data quality and sets a reference to the retrievability of higher quality transit spectra. 

The precisions for the synthetic data of HAT-P-1b assume median values on the precisions of the actual data for each instrumental mode. This renders different precisions for five instruments: HST STIS G430L (85 ppm), HST STIS G750L (140 ppm), HST WFC3 G141 (158 ppm), and two Spitzer band-passes at 3.6 (160 ppm) and 4.5 $\mu$m (141 ppm). In addition to the above precisions, the simulated data have been shifted with random Gaussian noise drawn from distributions with standard deviations matching the instrumental precisions to resemble genuine observations. The $p$--$T$ profile of the atmosphere is chosen such that the photospheric temperature at $\sim$1mbar is HAT-P-1b's equilibrium temperature of 1,320 K. The pressure associated with $R_p$ is $P_{\mr{ref}}=10^{-2}$ bar. The molecular and elemental abundances are representative of solar compositions at 1,320 K: log($X_{\mr{Na}}$) = $-5.5$, log($X_{\mr{K}}$) = $-6.7$, log($X_{\mr{H_2O}}$) = $-3.2$, log($X_{\mr{CH_4}}$) = $-4.4$, log($X_{\mr{CO}}$) = $-3.4$, log($X_{\mr{CO_2}}$) = $-6.9$, log($X_{\mr{HCN}}$) = $-7.8$, and log($X_{\mr{NH_3}}$) = $-5.9$ \citep{madhu12, asplund09, heng16}. The haze and cloud parameters are assumed to be $a = 10^{2}$, $\gamma = -7$, $P_{\mr{cloud}} = 10^{-2}$ bar, and $\bar{\phi} = 0.4$. The temperature of the photosphere is $T_{\mr{phot}}$ = 5,980 K and the stellar disk contains cool starspots at $T_{\mr{het}}$ = 5,100 K with a covering fraction of $\delta = 5 \%$. The high resolution forward models in the retrieval were evaluated using 3,000 wavelength points between 0.3 and 5 $\mu$m. The multi-dimensional space of forward model parameters is explored with 1,000 live points in the {\sc MultiNest} sampling algorithm. This amount of live points is an optimal via media in maximizing the accuracy of parameter estimates and the evidence and minimizing the time needed to attain a converged solution.

The priors used for the planetary and stellar parameters for all the retrievals in this work are displayed in Table \ref{tab:ret_priors}. The prior distributions and ranges follow similar prescriptions used in previous studies \citep[e.g.][]{benneke12,line12,macdonald17} and are based on physically plausible assumptions for the various parameters. A uniform prior distribution is used for parameters which vary less than 2 dex, while a uniform-in-logarithm prior is adopted for parameters whose values span over many orders of magnitude. A Gaussian prior distribution is used for the photospheric temperatures since they are rather reliably determined; parallaxes and angular sizes are used to derive a linear radius of the star, which together with a bolometric luminosity measurement gives the photospheric temperature from the Stefan-Boltzmann equation \citep[see e.g.][]{boyajian15}.

As for the prior ranges, the lower prior of the heterogeneity temperature derives from Figure 7 of \citet{berdyugina05} while the upper bound represents typical temperatures of solar faculae \citep{hirayama78}. The heterogeneity covering fraction $\delta$ spans from no coverage to a nominal upper limit based on maximal values consistent with amplitudes of stellar photometric variations \citep[see e.g.][]{rackham18}. The cloud and haze coverage fraction can span the whole range from clear (i.e. 0) to fully cloudy/hazy (i.e. 1). Since the atmosphere is structured from $10^2$ bar to 1 $\mu$bar, the cloud-top pressure can span the whole range, with smaller values potentially indicative of stronger vertical mixing. The lower limit of $\gamma$ is an extreme value indicative of hazes having imaginary refractive indices that vary strongly with wavelength \citep[e.g. see sulphides in][]{pinhas17} and of particle sizes $\lesssim$0.1 $\mu$m, while the upper bound of $\sim$0 represents larger particles. It can be shown that the Rayleigh-enhancement factor $a$ is $a = (X_{\mr{haze}}\sigma_{\mr{haze,0}})/(X_{\mr{H_2}}\sigma_0)$, where $X_{\mr{haze}}$ is the volumetric mixing ratio of hazes, $X_{\mr{H_2}}$ is the volumetric mixing ratio of molecular hydrogen, $\sigma_{\mr{haze,0}}$ is the cross-section of hazes at $\lambda_0$ (see Section 2.1.3), and $\sigma_0$ is the H$_2$ scattering cross-section at $\lambda_0$. \citet{pinhas17} use experimental refractive index data for a dozen aerosols and show their extinction cross-sections for small particles ($\sim$$10^{-2}$ $\mu$m) to be $\sim$$10^{-12}$\,${\mr{cm}}^{2}$ at $\lambda_0$. For nominal values of $X_{\mr{haze}}$ ranging from $10^{-20}$ to $10^{-8}$, a $X_{\mr{H_2}}$ of 0.85, and a $\sigma_0$ of $5.31\times10^{-27}$ $\mr{cm^2}$ (see Section 2.1.3), we find that $a$ spans from about $10^{-4}$ to $10^{8}$.

The molecular abundances $X_i$ range from extremely trace amounts ($10^{-12}$) to potentially metal-rich atmospheres ($\sim$1). The parametric $p-T$ profile we use is sectioned into three layers. The third layer is the regime where a high optical depth leads to an isothermal temperature profile and $P_3$ is the pressure defining the onset of this layer. Hot atmospheres can have substantial isothermal layers with $P_3$$\sim$$10^{-2}$ bar, with larger values possible for cooler atmospheres. Layer 1 and 2 are defined by $P_1$ and $P_2$; in the most general case, they can span the whole atmosphere but are conditional on being less than $P_3$. Gradients in the temperature profile are controlled by $\alpha_1$ and $\alpha_2$, which in principal can span very steep changes in temperature with height (i.e. values of $\sim$0) to minimal changes in $dT/dz$ (i.e. values of $\sim$1). The skin temperature at the top of the atmosphere spans cool temperatures of 800 K to slightly above the calculated equilibrium temperature.

Figure \ref{fig:simret} shows the retrieved posterior distributions of parameters for the simulated data and the retrieved spectral fit and $p$--$T$ profile are shown in Figure \ref{fig:fit_pT}. Both figures are made available on the Open Science Framework\footnote{\label{OSFlink}\url{https://osf.io/6gwtm/?view_only=46c6e5ece1264da598dc461948873055}} under `HAT-P-1b/Self-consistency test'. The parameter estimates are generally in good agreement with the true values used to generate the simulated observations. The inverted temperature profile is in excellent agreement with the true profile, with only marginal deviations between true and median profiles; the true $p$--$T$ parameter values all lie well within the retrieved 1$\sigma$ ranges. The majority of true molecular abundances are constrained within the retrieved 1$\sigma$ uncertainties, but overall are inverted less well than the $p$--$T$ parameters since the true abundances of CO and NH$_3$ lie just outside 1$\sigma$ estimates. The precisions and limited spectral coverage of the data inhibit sensitivity to the latter abundances. Importantly, the haze and cloud properties effective in the optical and near-infrared are constrained within the 1$\sigma$ confidence regions.

The stellar parameters are well constrained to the true inputs. There is a joint degeneracy between the covering fraction $\delta$ and the heterogeneity temperature T$_{\mr{het}}$. The positive correlation between these two stellar parameters arises from a decrease in the level of the optical transit depth with higher $T_{\mr{het}}$. Higher values of the heterogeneity temperature provide less contrast between the stellar photosphere and the heterogeneity features, reducing the transit depth in the 0.3 to 1.0 $\mu$m range as illustrated in the middle panel of Figure \ref{fig:het_effects}. Larger covering fractions are able to increase the transit depth and restore the spectrum to the same level. 

The simulated exercise demonstrates several points of essence. We illustrate that concurrent estimates of exoplanet atmosphere properties and features of a heterogeneous stellar disk are possible with present observations. In light of the dominant effect of stellar heterogeneity in the optical region, it is noteworthy that the presence and abundances of Na and K can be estimated reliably to 1$\sigma$. In general, the retrievability of stellar and planetary parameters must be examined individually for each considered planet. The ever increasing quality of forthcoming observations such as those from the JWST and future ground-based facilities as well as present instruments (e.g. VLT and ELTs) promises to increase the reliability of retrieved properties for transiting planets orbiting active stars.

\section{Application to Transmission Observations}\label{retrieval_application}

We here use our retrieval framework to study the stellar and planetary properties of available hot Jupiter observations. We apply our joint retrieval framework to the hot Jupiter ensemble presented in \citet{sing16} with the exception of HD 189733b, for which a transmission spectrum uncorrected for imprints of stellar heterogeneity is not available. Specifically, we focus on determining whether the spectra demonstrate imprints of stellar heterogeneity and clouds/hazes in the planetary atmosphere, with this information deriving especially from the 0.3 to 1.0 $\mu$m range. A study of the chemical compositions and abundances in the hot Jupiters are presented in a separate work (Pinhas et al., submitted).

Retrieval of the transmission spectra requires that the data have not been processed and corrected for stellar activity features. Therefore, we first outline the pre-processed data used as input to the retrieval framework. We then present our retrieval analyses on these pre-processed observations to illuminate the relative significance of stellar heterogeneity and clouds/hazes for the hot Jupiter ensemble.

\subsection{Data Uncorrected for Stellar Heterogeneity}\label{uncorrected_data}

A portion of the hot Jupiter transmission observations presented in \citet{sing16} have been processed and corrected for unocculted and occulted heterogeneity features. Therefore we remove any applied heterogeneity corrections to the spectra before a retrieval application is carried out on the ensemble. Table \ref{tab:system_properties} lists the data components as well as system properties for each planet. In what follows, we briefly describe the pre-processed data used for the planets.

\subsubsection{WASP-19b}

The uncorrected transmission spectrum of WASP-19b in the optical and near-infrared are from \citet{huitson13}. We obtain the uncorrected spectrum by reversing the correction to the four HST/STIS data points detailed by \citet{huitson13}. We note that \citet{sing16} present two more data points in the blue wavelength region that are not present in the optical data of \citet{huitson13}. However, sufficient information is not present in \citet{sing16} to determine the applied corrections to these points. We therefore do not use these two additional data and consider only the optical observations from \citet{huitson13}. In addition to these optical and near-infrared observations, we use two Spitzer data presented in \citet{sing16} to which no stellar correction was applied.

\subsubsection{WASP-6b}

To produce the WASP-6b transit spectrum uncorrected for heterogeneity of its star, we start with the data presented in \citet{nikolov15} and subtract the additional uncertainty applied in their Section 2.3. The typical uncertainty is now $\Delta R_p/R_{\star}$ $\sim$220 ppm smaller, in agreement with the change described in \citet{nikolov15}. This change is significant for most wavelength bins, representing a $\sim$40\% reduction in the error on the measurement.

\subsubsection{Other Hot Jupiters}

The transmission spectra for all other planets experienced no corrections. Therefore we use the data as presented in \citet{sing16} for HD 209458b, HAT-P-1b, WASP-39b, HAT-P-12b, WASP-31b, WASP-12b, and WASP-17b. Additional HST WFC3 data for WASP-39b \citep{tsiaras17}, HAT-P-12b \citep{line13}, and WASP-12b \citep{kreidberg14b} complement the above.

\subsection{Retrieval of Uncorrected Data}

\begin{table*}
\centering 
\begin{tabular}{c c c c c c c c}
\hline
\rule{0pt}{3ex}  Planet & log~$R^{'}_{\mr{HK}}$ & Spectral Type & $T_{\mr{phot}}$ (K) & $T_{\mr{eq}}$ (K) & $R_{p}$ ($R_J$) & Data & Ref.\\ [1ex]
\hline
\\ [-1.5ex]
\rule{0pt}{3ex}
 WASP-19b & $-4.660$ & G8 & $5500\pm100$ & 2,050 & 1.41 & \parbox{5cm}{ \centering HST/STIS/G430L, G750L \\ \centering HST/WFC3/G141 \\ \centering Spitzer/IRAC/3.6, 4.5 channel} & \parbox{2.3cm}{\centering \citet{huitson13} \\ \centering \citet{sing16}}\\[1ex]
 
 \\
 
  WASP-6b & $-4.741$ & G8 & $5375\pm65$ & 1,150 & 1.22 & \parbox{5cm}{ \centering HST/STIS/G430L, G750L \\ \centering Spitzer/IRAC/3.6, 4.5 channel} & \citet{nikolov15} \\[1ex]
 
 \\
 
  HD 209458b & $-4.970$ & G0 & $6092\pm103$ & 1,450 & 1.359 & \parbox{5cm}{ \centering HST/STIS/G430L, G750L \\ \centering HST/WFC3/G141 \\ \centering Spitzer/IRAC/3.6, 4.5 channel} &  \citet{sing16}\\[1ex]
 
 \\
 
  HAT-P-1b & $-4.984$ & G0 & $5980\pm8$ & 1,320 & 1.32 & \parbox{5cm}{ \centering HST/STIS/G430L, G750L \\ \centering HST/WFC3/G141 \\ \centering Spitzer/IRAC/3.6, 4.5 channel} & \citet{sing16}\\[1ex]
 
 \\
 
 WASP-39b & $-4.994$ & G8 & $5400\pm150$ & 1,120 & 1.27 & \parbox{5cm}{\centering HST/STIS/G430L, G750L\\ \centering HST/WFC3/G141 \\ \centering Spitzer/IRAC/3.6, 4.5 channel} & \parbox{2.3cm}{\centering \citet{tsiaras17} \\ \centering \citet{sing16}}\\[1ex]
 
 \\
 
  HAT-P-12b & $-5.104$ & K5 & $4650\pm60$ & 960 & 0.96 & \parbox{5cm}{ \centering HST/STIS/G430L, G750L \\ \centering HST/WFC3/G141\\ \centering Spitzer/IRAC/3.6, 4.5 channel} & \parbox{2.3cm}{\centering \citet{line13} \\ \centering \citet{sing16}}\\[1ex]

 \\
 
  WASP-31b & $-5.225$ & F & $6300\pm100$ & 1,580 & 1.55 & \parbox{5cm}{ \centering HST/STIS/G430L, G750L\\ \centering HST/WFC3/G141 \\ \centering Spitzer/IRAC/3.6, 4.5 channel} & \citet{sing16}\\[1ex]
 
 \\
 
  WASP-12b & $-5.500$ & G0 & $6400\pm190$ & 2,510 & 1.73 & \parbox{5cm}{ \centering HST/STIS/G430L, G750L \\ \centering HST/WFC3/G102, G141 \\ \centering Spitzer/IRAC/3.6, 4.5 channel} & \parbox{2.3cm}{\centering \citet{kreidberg15} \\ \centering \citet{sing16}}\\[1ex]
 
 \\
 
 WASP-17b & $-5.531$ & F4 & $6650\pm80$ & 1,740 & 1.89 & \parbox{5cm}{ \centering HST/STIS/G430L, G750L \\ \centering HST/WFC3/G141 \\ \centering Spitzer/IRAC/3.6, 4.5 channel} & \citet{sing16}\\[1ex]
 \\[-1.5ex]
\hline
\end{tabular}
\caption{Planetary system properties and observations. Columns two to four list properties of the stellar host and columns five to seven show planetary properties. The observations used for each planet and uncorrected for stellar activity effects are shown in the eighth column, with relevant references for the uncorrected data in the last column.}
\label{tab:system_properties}
\end{table*}

\begin{table*}
\begin{center}
\centering
\begin{tabular}{|c|c|c|c|}
\hline
\rowcolor{HEADER}
Group & Planet & Stellar Heterogeneity & Clouds/Hazes \\ \hline
\rowcolor{I}
\multicolumn{1}{ |c| }{}  & WASP-6b & Strong & Weak - Substantial \\ 
\rowcolor{I}
\multicolumn{1}{ |c| }{\multirow{-2}{*}{I} }  & WASP-39b & Substantial & Weak - Strong \\ \cline{1-4}
\rowcolor{II}
\multicolumn{1}{ |c| }{} & HD 209458b & Weak & Substantial - Very Strong \\ 
\rowcolor{II}
\multicolumn{1}{ |c|  }{\multirow{-2}{*}{II}} & HAT-P-12b & Weak & Weak (against) - Very Strong \\ 
\cline{1-4}
\rowcolor{III}
\multicolumn{1}{ |c|  }{} & HAT-P-1b & Weak (against) & Weak (against) - Substantial \\ 
\rowcolor{III}
\multicolumn{1}{ |c|  }{\multirow{-2}{*}{III}} & WASP-31b & Weak (against) & Weak \\ \cline{1-4}
\rowcolor{IV}
\multicolumn{1}{ |c|  }{}  & WASP-19b & Substantial (against) & Weak - Substantial \\ 
\rowcolor{IV}
\multicolumn{1}{ |c|  }{}  & WASP-17b & Substantial (against) & Weak (against) \\ 
\rowcolor{IV}
\multicolumn{1}{ |c|  }{\multirow{-3}{*}{IV} }  & WASP-12b & Substantial (against) & Weak (against) - Weak \\ \cline{1-4}
\end{tabular}
\end{center}
\caption{Roles of stellar heterogeneity and clouds/hazes in the spectral ensemble. The qualitative descriptions in the two categories are based on the Bayes factor results listed in Table \ref{tab:model_comparisons} and the Bayes factor nomenclature scale of \citet{kass95}. Instances of `(against)' following the qualitative descriptions signify that the category is not favoured to the degree of the description. Thus, for example, the spectrum of WASP-31b is best interpreted through a weak suggestion against stellar heterogeneity and weak evidence for hazes and/or clouds. The results for WASP-6b represent the full set of data. In summary, four groups of planets are distinguished through the role in which stellar heterogeneity and clouds/hazes explain their spectra. Group I is best characterised by imprints of stellar heterogeneity and clouds/hazes. Group II comprises HD 209458b and HAT-P-12b and shows weak evidence for stellar heterogeneity but beyond substantial suggestions of clouds and/or hazes. HAT-P-1b and WASP-31b constitute Group III and show weak evidence against stellar heterogeneity but weak to substantial indications of clouds and/or hazes. The fourth group -- WASP-19b, WASP-17b, and WASP-12b -- can be explained best without stellar heterogeneity and weak to no evidence for clouds/hazes.
}
\label{tab:condensed_expl}
\end{table*}

\begin{figure*}
\centering
\includegraphics[scale=0.8]{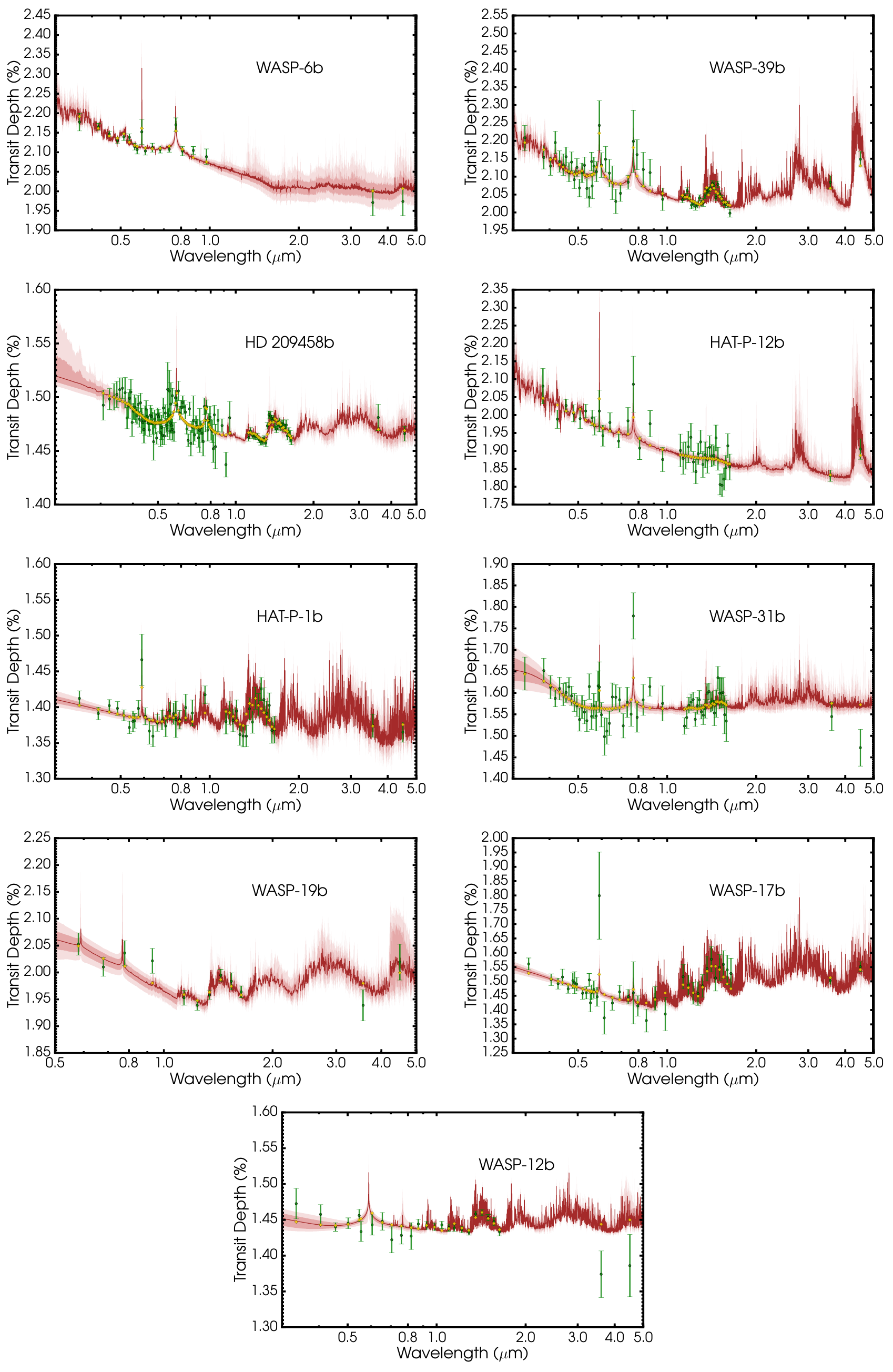}%
\caption{Retrieved model transmission spectra compared to observations for the ensemble of hot Jupiters. The data are shown in green and the retrieved median model is in dark red with associated 1$\sigma$ and 2$\sigma$ confidence contours. The yellow diamonds are the binned median model at the same resolution as the observations. Model type HMch is shown for WASP-6b, WASP-39b, and HD 209458b; model type HMc is shown for HAT-P-12b; model type Mch is shown for HAT-P-1b, WASP-31b, and WASP-19b; and model type M is shown for WASP-17b and WASP-12b.}
\label{fig:spectral_fits}
\end{figure*}

\begin{table*}
\caption{Bayesian model comparisons of the hot Jupiter ensemble. Listed are the evaluated model types, natural logarithms of the evidences $\mathcal{Z}$ and Bayes factors with their associated significance levels. The reference model (`Ref.') includes a formalism for the heterogeneous of the stellar disk, opacity contributions from H$_2$O, CH$_4$, CO, CO$_2$, HCN, NH$_3$, Na, K, H$_2$-H$_2$ and H$_2$-He, and a cloud and haze parameterisation. The Bayes factor $\mathcal{B}_{0i}$ represents the preference for the reference model over model $i$. The detection significance signifies the degree of inclination of the reference model over the alternative model. Detection significance calculations are valid for Bayes factors above one, hence the `N/A' values.}
\label{tab:model_comparisons}
\begin{center}
\centering
\begin{tabular}{ c c c c }
    \hline\hline
    \\[-2ex]
    \multicolumn{4}{c}{\textbf{WASP-6b (Full Data)}}\\[1ex]
    Model Type & Evidence (ln($\mathcal{Z}$)) & Bayes Factor $\mathcal{B}_{0i}$ & Detection Sig.\\[1ex]
    \hline
    \\[-1ex]
    HMch & 134.25 & Ref. & Ref.\\[1ex] 
    HMc & 134.06 & 1.22 & 1.37$\sigma$ \\[1ex] 
    HM & 132.39 & 6.48 & 2.48$\sigma$ \\[1ex] 
    Mch & 130.76 & 32.94 & 3.13$\sigma$ \\[1ex] 
    M & 119.20 & 3.46 $\times 10^{6}$ & 5.83$\sigma$\\[1ex]
    \hline\hline
    \\
    \hline\hline
    \\[-2ex]
    \multicolumn{4}{c}{\textbf{WASP-6b (Without Spitzer Data)}}\\[1ex]
    Model Type & Evidence (ln($\mathcal{Z}$)) & Bayes Factor $\mathcal{B}_{0i}$ & Detection Sig.\\[1ex]
    \hline
    \\[-1ex]
    HMch & 124.16 & Ref & Ref. \\[1ex] 
    HMc & 124.28 & 0.89 & N/A \\[1ex] 
    HM & 123.79 & 1.45 & 1.56$\sigma$ \\[1ex] 
    Mch & 124.06 & 1.11 & $1.23\sigma$ \\[1ex] 
    M & 124.47 & 0.74 & N/A \\[1ex]
    \hline\hline
    \\
    \hline\hline
    \\[-2ex]
    \multicolumn{4}{c}{\textbf{WASP-39b}}\\[1ex]
    Model Type & Evidence (ln($\mathcal{Z}$)) & Bayes Factor $\mathcal{B}_{0i}$ & Detection Sig.\\[1ex]
    \hline
    \\[-1ex]
    HMch & 401.44 & Ref. & Ref. \\[1ex] 
    HMc & 401.17 & 1.30 & $1.45\sigma$ \\[1ex] 
    HM & 394.17 & 1430 & 4.22$\sigma$ \\[1ex] 
    Mch & 399.60 & 6.26 & 2.47$\sigma$ \\[1ex] 
    M & 393.16 & 3923 & 4.47$\sigma$\\[1ex]
    \hline\hline
    \\
    \hline\hline
    \\[-2ex]
    \multicolumn{4}{c}{\textbf{HD 209458b}}\\[1ex]
    Model Type & Evidence (ln($\mathcal{Z}$)) & Bayes Factor $\mathcal{B}_{0i}$ & Detection Sig.\\[1ex]
    \hline
    \\[-1ex]
    HMch & 957.83 & Ref. & Ref. \\[1ex] 
    HMc & 956.64 & 3.31 & 2.14$\sigma$\\[1ex] 
    HM & 945.45 & 2.38$\times 10^{5}$ & 5.34$\sigma$ \\[1ex] 
    Mch & 957.47 & 1.44 & $1.55\sigma$\\[1ex] 
    M & 947.01 & 5$\times 10^{4}$ & 5.02$\sigma$ \\[1ex]
    \hline\hline
    \\
    \hline\hline
    \\[-2ex]
    \multicolumn{4}{c}{\textbf{HAT-P-12b}}\\[1ex]
    Model Type & Evidence (ln($\mathcal{Z}$)) & Bayes Factor $\mathcal{B}_{0i}$ & Detection Sig.\\[1ex]
    \hline
    \\[-1ex]
    HMch & 264.74 & Ref. & Ref. \\[1ex] 
    HMc & 264.83 & 0.92 &  N/A\\[1ex] 
    HM & 245.84 & 1.62$\times 10^{8}$ & 6.47$\sigma$ \\[1ex] 
    Mch & 264.56 & 1.20 & $1.35\sigma$ \\[1ex] 
    M & 245.67 & 1.9$\times 10^{8}$ & $6.50\sigma$ \\[1ex]
    \hline\hline
\end{tabular}
\end{center}
\end{table*}

\begin{table*}
\begin{center}
\begin{tabular}{ c c c c }
    \hline\hline
    \\[-2ex]
    \multicolumn{4}{c}{\textbf{HAT-P-1b}}\\[1ex]
    Model Type & Evidence (ln($\mathcal{Z}$)) & Bayes Factor $\mathcal{B}_{0i}$ & Detection Sig.\\[1ex]
    \hline
    \\[-1ex]
    HMch & 302.16 & Ref. & Ref. \\[1ex] 
    HMc & 302.23 & 0.93 & N/A\\[1ex] 
    HM & 299.98 & 8.90 &  $2.62\sigma$\\[1ex] 
    Mch & 302.57 & 0.69 & N/A\\[1ex] 
    M & 296.07 & 443.86 & $3.92\sigma$ \\[1ex]
    \hline\hline
    \\
    \hline\hline
    \\[-2ex]
    \multicolumn{4}{c}{\textbf{WASP-31b}}\\[1ex]
    Model Type & Evidence (ln($\mathcal{Z}$)) & Bayes Factor $\mathcal{B}_{0i}$ & Detection Sig.\\[1ex]
    \hline
    \\[-1ex]
    HMch & 395.54 & Ref. & Ref. \\[1ex] 
    HMc & 394.98 & 1.75 & $1.72\sigma$ \\[1ex] 
    HM & 395.17 & 1.45 & $1.56\sigma$ \\[1ex] 
    Mch & 395.98 & 0.64 & N/A \\[1ex] 
    M & 392.00 & 34.51 & 3.14$\sigma$ \\[1ex]
    \hline\hline
    \\
    \hline\hline
    \\[-2ex]
    \multicolumn{4}{c}{\textbf{WASP-19b}}\\[1ex]
    Model Type & Evidence (ln($\mathcal{Z}$)) & Bayes Factor $\mathcal{B}_{0i}$ & Detection Sig.\\[1ex]
    \hline
    \\[-1ex]
    HMch & 76.00 & Ref. & Ref. \\[1ex] 
    HMc & 74.89 & 3.06 & 2.09$\sigma$ \\[1ex]
    HM & 75.71 & 1.34 & $1.48\sigma$ \\[1ex] 
    Mch & 77.27 & 0.28 & N/A \\[1ex] 
    M & 72.59 & 30.53 & $3.10\sigma$ \\[1ex]
    \hline\hline
    \\
    \hline\hline
    \\[-2ex]
    \multicolumn{4}{c}{\textbf{WASP-17b}}\\[1ex]
    Model Type & Evidence (ln($\mathcal{Z}$)) & Bayes Factor $\mathcal{B}_{0i}$ & Detection Sig.\\[1ex]
    \hline
    \\[-1ex]
    HMch & 249.31 & Ref. & Ref. \\[1ex] 
    HMc & 249.98 & 0.51 & N/A \\[1ex] 
    HM & 250.08 & 0.47 & N/A \\[1ex] 
    Mch & 250.52 & 0.30 & N/A \\[1ex] 
    M & 251.02 & 0.18 & N/A \\[1ex]
    \hline\hline
    \\
    \hline\hline
    \\[-2ex]
    \multicolumn{4}{c}{\textbf{WASP-12b}}\\[1ex]
    Model Type & Evidence (ln($\mathcal{Z}$)) & Bayes Factor $\mathcal{B}_{0i}$ & Detection Sig.\\[1ex]
    \hline
    \\[-1ex]
    HMch & 195.33 & Ref. & Ref. \\[1ex] 
    HMc & 196.38 & 0.35 & N/A \\[1ex] 
    HM & 195.20 & 1.14 & 1.28$\sigma$ \\[1ex] 
    Mch & 197.34 & 0.13 & N/A \\[1ex] 
    M & 196.79 & 0.23 & N/A \\[1ex]
    \hline\hline
\end{tabular}
\end{center}
\end{table*}

We conduct five independent retrievals for each hot Jupiter to examine the roles of stellar heterogeneity and clouds/hazes contained in the spectral ensemble. The roles of stellar heterogeneity and clouds/hazes in the hot Jupiter spectra are summarised in Table \ref{tab:condensed_expl}. Table \ref{tab:model_comparisons} provides the detailed statistical results for the five retrieval types used to arrive at the conclusions in Table \ref{tab:condensed_expl}. The full retrieval results including posterior distributions, spectral fits, and $p$--$T$ profiles are available on the Open Science Framework\textsuperscript{\ref{OSFlink}}. The retrieved stellar parameters of the most complex model described below (i.e. number 1) are listed in Table \ref{tab:stelparams}. Representative spectral fits for the planetary ensemble are displayed in Figure \ref{fig:spectral_fits}. All retrieval types include parameters of the $p$--$T$ profile and mixing ratios of eight molecules and alkali metals and are distinct as follows:
\begin{enumerate}[1.]
    \item \textbf{Heterogeneity + Molecules + Clouds/Hazes (hereafter, HMch)}: Three stellar parameters, eight molecular and elemental abundances, and four cloud and haze parameters (grey and non-grey opacities, respectively). 
    \item \textbf{Heterogeneity + Molecules + Clouds (hereafter, HMc)}: Three stellar parameters, eight molecular and elemental abundances, and two cloud parameters (grey opacity only). This model excludes non-grey opacity from hazes.
    \item \textbf{Heterogeneity + Molecules (hereafter, HM)}: Three stellar parameters and eight molecular and elemental abundances. This model excludes haze and cloud opacities.
    \item \textbf{Molecules + Clouds/Hazes (hereafter, Mch)}: Eight molecular and elemental abundances and four cloud and haze parameters (both non-grey and grey opacities). This model excludes stellar heterogeneity. 
    \item \textbf{Molecules (hereafter, M)}: Eight molecular and elemental abundances. This model excludes both stellar heterogeneity and clouds/hazes.
\end{enumerate}

The retrieval analyses reveal four groups of planets distinguished through the role in which stellar heterogeneity and clouds/hazes explain the spectra. This is summarised in Table \ref{tab:condensed_expl}. The first group comprises WASP-39b and WASP-6b. Their spectra are best characterised by imprints of stellar heterogeneity and clouds and/or hazes. HD 209458b and HAT-P-12b comprise the second group for which there is high importance of hazes and/or clouds and weak evidence for stellar heterogeneity. The third group constitutes HAT-P-1b and WASP-31b and shows weak evidence against stellar heterogeneity but weak to substantial indications of clouds and/or hazes. The fourth group -- WASP-19b, WASP-17b, and WASP-12b -- are fit best by alkali and molecular absorbers with H$_2$ scattering without stellar heterogeneity and weak to no evidence for clouds/hazes. We here discuss the detailed role of stellar heterogeneity and clouds/hazes for planets in each group.

\subsubsection{Group I: Stellar Heterogeneity + Clouds/Hazes}

Table \ref{tab:condensed_expl} shows the spectra of WASP-6b and WASP-39b are best explained in light of stellar heterogeneity as well as clouds and hazes.

The retrieval results for the full dataset of WASP-6b are shown in Table \ref{tab:model_comparisons}. The full observations of WASP-6b show the greatest suggestion for stellar heterogeneity imprints among the planetary sample. The significance of stellar heterogeneity in the spectrum is 3.13$\sigma$ through a comparison of the evidences for HMch and Mch model types. The Bayes factor of HMch to HMc models is 1.22 ($1.37\sigma$), indicating weak evidence for hazes in addition to stellar heterogeneity. The combined evidence for hazes and clouds is 2.48$\sigma$ (Bayes factor of 6.48) and is thus substantial. Therefore models with heterogeneity are favoured strongly with hazes increasing the fit only marginally. Nevertheless, stellar heterogeneity and planetary hazes/clouds are complementary in best interpreting the full spectrum. The retrieved results of the HMch model suggest an active star with a covering fraction of $12.1^{+2.2}_{-1.9}\%$. This substantial active area is dominated by cool spots with temperatures of $4556^{+242}_{-213}$ K, indicating large temperature contrasts of $\sim$800 K between the photosphere ($\sim$5375 K) and regions of unocculted cool spots. This evidence for widespread unocculted cool spots is intriguing as photometric monitoring suggests that the flux of the star does not vary by more than 1$\%$ in V band \citep{nikolov15}. However, photometric monitoring places only a lower limit on the heterogeneity, as longitudinally symmetric active regions do not contribute to the observed variability \citep{rackham18}. Furthermore, the Ca II H and K stellar activity index for WASP-6, log $R'_{\mr{HK}}$ = $-4.741$, which is higher than all in this sample except WASP-19, points to a high level of chromospheric activity, which in turn correlates to higher photospheric activity \citep{lockwood07}.

To investigate the influence of the Spitzer data on the above suggestions, we have carried out another set of five retrievals which includes one of each model type described above but with the Spitzer data excluded. In this case, a heterogeneity explanation is not statistically strong since the Bayes factor of HMch to Mch models is 1.28 ($1.43\sigma$). Moreover, the STIS data indicate weak evidence against hazes (Bayes factor of 0.89). The existence of grey opacity clouds is favoured in both scenarios given the increase in significances between HMc and HM models. In all, it is clear that a definitive (non)detection of heterogeneity from the transmission spectrum of WASP-6b will need to await future high quality observations, specifically high resolution and high precision observations in the 0.3 $-$ 1.0 $\mu$m range where heterogeneity effects are most pronounced. Observations in the HST WFC3 bandpass could also detail the degree to which molecular absorption features from the planetary atmosphere are muted by the presence of clouds/hazes or enhanced by cool star spots.

The full observations of WASP-39b show a second suggestion of stellar heterogeneity imprinted in a transmission spectrum. The Bayes factor of HMch to Mch models is 6.26, indicative of a substantial heterogeneity signal at 2.47$\sigma$. The retrieved covering fraction of heterogeneity features is $10.3^{+5.9}_{-3.7}\%$. The coverage area is dominated by cool spots with temperatures of $4936^{+240}_{-309}$ K and photospheric temperature contrasts of $\sim$500 K. The general presence of clouds and hazes is evident at 4.22$\sigma$ (Bayes factor of 1430) and the spectral component arising from hazes is suggested to be weak at 1.45$\sigma$ (Bayes factor of 1.30).

\subsubsection{Group II: Weak Stellar Heterogeneity + Clouds/Hazes}

Table \ref{tab:condensed_expl} shows the spectra of HD 209458b and  HAT-P-12b are best explained in light of hazes and/or clouds with weak evidence for stellar heterogeneity. HD 209458b and HAT-P-12b display slender hints of heterogeneity considering Bayes factors of 1.44 and 1.20 (respectively, 1.55$\sigma$ and 1.35$\sigma$) between HMch and Mch models. The retrieved covering fraction of heterogeneity features for HD 209458 (HAT-P-12) is  $3^{+2}_{-1}\%$ ($18^{+3}_{-2}\%$). Cool spots with temperatures of $4167^{+119}_{-157}$ K dominate the coverage area for HAT-P-12, giving a photosphere-spot temperature contrast of $\sim$$500$ K. On the other hand, the spots on HD 209458 have temperatures of $\sim$3500 K and/or $\sim$5700 K. The general presence of clouds and hazes in HD 209458b is evident at 5.34$\sigma$ (Bayes factor of 2.4$\times10^{5}$) and the spectral component arising from hazes is suggested to be substantial at 2.14$\sigma$ (Bayes factor of 3.31). In the case of HAT-P-12b, a Bayesian interpretation weakly favours a model without hazes since the Bayes factor of HMch to HMc models is 0.92 but illustrates a very strong suggestion for clouds at more than $6\sigma$.

\subsubsection{Group III: Weak Evidence Against Stellar Heterogeneity + Weak to No Clouds/Hazes}

Table \ref{tab:condensed_expl} shows the spectra of HAT-P-1b and WASP-31b are best explained through a weak suggestion against stellar heterogeneity and weak to no evidence for clouds/hazes. HAT-P-1b and WASP-31b constitute spectral interpretations with a weak suggestion against a heterogeneity component since the Bayes factors of HMch to Mch models are below one. The Bayes factors in Table \ref{tab:model_comparisons} imply the importance of clouds over hazes in the spectrum of HAT-P-1b, while hazes are weakly favoured in WASP-31b.

\subsubsection{Group IV: Substantial Evidence Against Stellar Heterogeneity + Weak to No Clouds/Hazes}

Table \ref{tab:condensed_expl} shows the spectra of three planets -- WASP-19b, WASP-17b, and WASP-12b -- are best explained through substantial evidence against stellar heterogeneity and weak or no evidence for cloud/haze coverage. In the case of WASP-19b, the model with the largest Bayesian evidence is Mch, such that stellar heterogeneity is not substantially preferred. Clouds and hazes are suggested to a weak degree through comparison of models HMc and HM with HMch. On the other hand, extant features in the spectra of WASP-17b and WASP-12b are sufficiently explained through molecular and alkali absorption as well as H$_2$ Rayleigh scattering, i.e. model type M, provided that its Bayes factor is generally less than other model realisations.

The result for WASP-19b is surprising as WASP-19 displays the highest log $R'_{\mr{HK}}$ index of the host stars in the current sample (see Table \ref{tab:system_properties}), suggesting a high level of chromospheric activity. Corresponding photospheric activity has been observed in the form of V-band rotational variability \citep{huitson13} and spot-crossing events during transit \citep{mancini13, huitson13,sedaghati18}. However, the stellar contribution to transmission spectra depends on a variety of parameters, including active region temperature contrasts, covering fractions, and locations within the projected stellar disk at the time of the observation. Nevertheless, we have explored whether the unexpected lack of evidence for stellar contamination for WASP-19b is a result of the two-component stellar heterogeneity model of Equation (\ref{observed_TD}) which assumes the prominence of either cool spots or hot faculae. We have conducted two additional retrievals that use a modified model that treats spots and faculae as separate components following Equation (3) of \citet{rackham18}. The two retrievals are made available on the Open Science Framework\textsuperscript{\ref{OSFlink}} under `WASP-19b/Three-component photosphere'. In the case of the HMch model with spots and faculae treated independently, we find that the Bayes factor of Mch to HMch models is above 3. This lack of evidence for heterogeneity is also supported by the fact that the retrieved covering fractions of faculae and spots in the HMch model are consistent with zero. Clouds and hazes are suggested at 2.5$\sigma$ (Bayes factor of 6.75) by comparing the HMch and HM models and the cloud/haze properties are also not significantly different than those of the two-component heterogeneity model. The results from a three-component heterogeneity model validate the suggestion that the WASP-19b data show a lack of evidence for stellar heterogeneity.

The case of WASP-19b illustrates that the relationship between stellar activity and contamination of transmission spectra might not correlate in all cases. Still, given that WASP-19b possesses the least constraining optical spectrum of the current sample, consisting of only four points between 0.5 and 1.0 $\mu$m,  higher-resolution optical observations uncorrected for stellar effects can elucidate this relationship further. We emphasize that the suggestion against stellar heterogeneity for WASP-19b might be driven by the low data quality of the available spectrum in addition to multiple limiting facets discussed in Section \ref{sec:limitations}.

\section{Atmospheric-Photospheric Retrieval: Limitations and Future Steps}\label{sec:limitations}

Our joint retrieval methodology and its first applications provide a novel insight into how stellar contamination can affect some of the published transmission spectra. It also shows encouraging results in disentangling stellar contamination from genuine planetary signatures. We here briefly explore the key assumptions and  limitations of our methodology as well as constructive future developments in joint retrievals for better elucidation of stellar and exoplanetary properties.

\subsection{Model Components: Hazes and Starspots}

Our Bayesian approach is based on the assumptions that the model components used are {\em complete} (i.e., no significant component is missing) and {\em correct} (i.e., the models and its components rightly capture the key properties of the real system). Although our model components and therefore these underlying assumptions are supported by state-of-the-art knowledge on exoplanets and are also widely used, it is important to point out that they are still assumptions. It is thus prudent to briefly review the uncertainties implicit to our assumptions and, especially, place them in the context of stellar contamination, which is explored in this study.

It is important to realize that we probably do not have a complete and correct model for stellar contamination yet:  stellar heterogeneity (the spatial distribution of temperature/spectral variations across stellar disks and its temporal evolution as a function of stellar parameters)  is very poorly understood (for recent summaries see, e.g., \citealt{rackham18} and \citealt{Apai2018}). For now, we do not have a good handle on basic parameters such as starspot covering fractions and typical temperature distributions of spots and faculae in stars other than the Sun.

The stellar contamination model chosen in this study is state-of-the-art; nevertheless, the spectral components that we use for the heterogeneities are simply stellar photospheric model spectra, not starspot/facular spectra \citep[e.g.,][]{norris17}. As of now, no dataset exists to allow us to test the validity of our stellar heterogeneity predictions on explicit starspot/facular spectra to the level that would be desirable; future efforts should be directed to build such datasets from stellar and heliophysical observations.

The limitations of our model for stellar contamination are especially significant due to its potential to resemble hazes, which have been invoked in exoplanetary atmospheres and also included in our models. We note that these unidentified particles are represented by featureless spectra, where the extinction coefficients of the assumed particles is free to vary by four or more orders of magnitude relative to that of H$_2$ scattering. As the putative haze particles do not have any other detectable features in the optical, their existence is only inferred from residual spectral slopes that are not otherwise explained by planetary atmosphere models.

However, as recognized in the literature \citep{mccullough14,rackham17} and also shown by our results (see Table \ref{tab:condensed_expl}), some transit spectra can be better reproduced by stellar contamination as well. The distinction of hazes and stellar contamination on the basis of the Bayesian evidence is necessarily based on the assumption that we have complete and correct models for both: future efforts should focus on testing and verifying these assumptions, as hazes and stellar contamination may have degenerate spectral signatures.

\subsection{Stellar Variability}

Our methodology represents an important step forward in disentangling stellar contamination and genuine planetary features but still only focuses on the transmission data. It does not yet incorporate fully the evidence provided by focused stellar variability observations. Stellar heterogeneity can be, to some level, constrained by stellar variability measurements \cite[e.g.,][]{sing11, aigrain12, zellem17} and by absorption and emission line measurements \citep[e.g.,][]{boisse09}. It is important to note that neither these nor other methods yet allow precise measurements of spot/facular covering fractions or temperature distributions over the stellar disk; variability measurements can only place a lower limit on these parameters \citep{jackson13,rackham18}.

Perhaps the most important information provided by stellar activity indicators is that stellar disks are heterogeneous and that stellar contamination, to some level, must occur. Therefore, in principle, uniform priors on the models (as in the present work) should not be assumed, though the weight that should be assigned to models is challenging to quantify in practice. Future studies should attempt to integrate an improved understanding of an individual stellar disk's heterogeneity (spatial distribution of regions with unique temperatures/spectra) through priors informed by the combined knowledge from spectroscopic and photometric activity indicators.

\subsection{Multi-epoch Observations}

Due to the limited wavelength range of any single spectrograph or imager, most exoplanetary transmission spectra have been obtained with multiple instruments and at multiple epochs, often separated by months or years. Stellar heterogeneity, however, varies in time. Therefore, stellar contamination imprinted in a spectrum can vary between epochs. By necessity our study has analysed the planet transit and stellar contamination signals as if the observations occurred in a single epoch. A multi-epoch study is currently beyond the information content of the available datasets; nevertheless, future observations should ensure that data are collected in a fashion that facilitates retrieval of multi-epoch data. Stellar contamination, if significant, will be time-variable. Assuming an average, representative contamination (as in our study) will likely not be satisfactory for objects where some datasets have been collected at times of greater stellar activity.

\subsection{Analysis Restricted to Lower-activity Stars}

It is important to note that our current methodology naturally excludes the exoplanets for which the most significant stellar contamination is expected. Some high-quality datasets published could not be included for the reason that stellar activity corrections were so significant that the data could not be reasonably `un-corrected' and then represented with a single-epoch stellar contamination model.

A good example for this limitation is HD~189733b for which HST datasets using multiple STIS, ACS, and WFC3 modes have been published, representing data collected over 22 different epochs. The host star HD~189733 itself is clearly active: significant photometric variations at the 3\% level have been observed between different epochs \citep{sing11}.

\citet{pont13} and subsequent studies using these data are based on a stellar contamination correction that applies an individual, multiplicative, and wavelength-dependent correction (blackbody slope) to each dataset. The correction is derived from an estimated starspot covering fraction (determined for each epoch) and a starspot temperature (constant in time, and derived on the basis of a single starspot occultation by HD~189733b).

Although the \citet{pont13} study certainly represented a state-of-the-art analysis, the stellar correction method used was based on optimistic assumptions and had four significant limitations, several of which were also acknowledged in that work (see their Section 3.3). As these and similar assumptions are often used in the literature to combine multi-epoch data, it is instructive to review their limitations.  First, the systematic uncertainties introduced by the starspot/activity corrections have not been analyzed fully or folded into the uncertainties on the data. Second, the corrections optimistically assume the lowest possible starspot covering fraction and stellar contamination, because the starspot areal covering fractions are derived from photometric variability assuming a linear relation. This linear relation is valid only in the most optimistic cases, as pointed out by \citet{pont13}, a finding later quantified independently for general cases by \citet{jackson13} and \citet{rackham18}. Third, the spectral slope of the stellar contamination correction applied by \cite{pont13} is constrained by spot-crossing events collected for very large spots. It is likely that due to strong observational selection biases that very large spots are not representative of the size population of starspots/faculae that exist on HD~189733. Fourth, no photometric data existed for some of the epochs in which data were taken and \citet{pont13} were forced to interpolate starspot covering fractions between neighboring epochs, which is unlikely to provide a correct value. That study also uses a common assumption that planetary transmission spectra must be continuous across different instruments and applies an offset to ensure this. However, this step, intended to compensate for stellar brightness variations, does not account for the color terms intrinsic to the stellar heterogeneity.

The HD~189733b dataset -- used in a significant number of follow-up studies -- demonstrates the challenges in correcting for stellar contamination in moderately active stars and highlights untested -- and in retrospect rather optimistic -- assumptions that underpin the corrections that are introducing poorly understood and possibly significant systematic errors in such datasets.

Our study did not include data from significantly active stars whose stellar luminosities have changed by more than a few per cent between epochs, and thus our sample of planets likely does not represent the full range of stellar contamination that should be expected in transit data.

\section{Discussion and Conclusions}\label{discussion_conclusions}

The four groups of planets identified in Table \ref{tab:condensed_expl} are approximately arranged by increasing chromospheric activity index (log $R'_{\mr{HK}}$), as seen by comparison of Tables \ref{tab:system_properties} and \ref{tab:condensed_expl}. The evident exceptions to this arrangement are WASP-19b and WASP-39b. Highly active stars (i.e., WASP-6) are found in the first group with evidence for stellar heterogeneity while low activity stars (i.e. WASP-12 and WASP-17) settle into the fourth group with substantial evidence against heterogeneity. The remaining planets (excepting WASP-19b and WASP-39b) are essentially ordered into the second and third groups, with weak evidence for or against heterogeneity. The observational consequences of this rough ordering of planets based on log $R'_{\mr{HK}}$ is noteworthy, for it suggests that the popular chromospheric activity indicator offers some predictive power as to whether an exoplanet transmission spectrum will be affected by stellar heterogeneity.

The substantial and yet indefinite suggestions of stellar heterogeneity in the spectra of WASP-6b and WASP-39b offer definite observational strategies to increase the significance of inferred stellar heterogeneity effects and planetary properties with our retrieval methodology. While the effects of stellar heterogeneity can be significant at wavelengths as long as 2 $\mu$m (see Figure \ref{fig:het_effects}), they are most pronounced between 0.3 and 1 $\mu$m and thus high-impact observations should focus on this spectral range to probe the activity of stellar photospheres. The corollary is that infrared observations essentially ascertain atmospheric properties of the exoplanet alone.

Joint studies of exoplanetary atmosphere compositions and heterogeneous stellar photospheres will benefit from precision observations in the Na and K absorption bands achievable with multiple HST orbits using STIS 430L and 750L grisms. Cool spots can masquerade as Na and K absorption features for a variety of stellar temperatures, metallicities, and gravities. In such cases and for observations of limited precisions (i.e. $\lesssim$100 ppm), degenerate explanations arise between Na and K abundances and stellar activity influence. Higher-precision observations in the Na and K bands can break this degeneracy, in turn enabling more precise estimates of heterogeneity covering fractions, alkali abundances, and haze properties. In addition, joint inferences will also benefit from high-resolution, broadband observations in the 0.3 to 1.0 $\mu$m window. This aspiration is already achievable through intelligent use of the VLT FORS2 spectrograph \citep[see for e.g.][]{sedaghati18}. Such finely-sampled observations can serve as a microscope into the heterogeneous coverage fraction since $\delta$ amplifies features of the heterogeneous spectrum (see Figure \ref{fig:het_effects}, upper panel). The positive correlation between the coverage fraction and heterogeneity temperature seen in Figure \ref{fig:simret} would also imply better constraints on the heterogeneity temperature  with such VLT observations.

The JWST can also contribute much to the joint retrieval of stellar and exoplanetary properties. The multi-wavelength capabilities facilitated by JWST can help separate stellar and planetary spectral imprints as the contrast between heterogeneous zones and the pristine stellar photosphere decreases with longer wavelengths as shown in Figures \ref{fig:stel_components}-\ref{fig:het_effects}. Thus, for example, spectra across 5 to 28 $\mu$m obtained with observing modes of the MIRI instrument will essentially reflect properties of the exoplanetary atmosphere alone. Simultaneous study of stellar and exoplanetary properties will also be possible. The time-series observing mode of the NIRCam instrument will permit stellar activity monitoring in the 0.6 to 5.0 $\mu$m range. The amplitude of time-series photometry can reveal lower limits on heterogeneous covering fractions \citep{rackham18} which can in turn be used as retrieval constraints using spectroscopy in the 0.6 to 5.0 $\mu$m range from NIRISS and NIRSpec modes.

In summary, we have presented a new retrieval methodology, {\sc Aura}, that enables simultaneous inference of the properties of exoplanetary atmospheres and their host stars in the transiting configuration. The developed framework permits the inference of general in-homogeneous properties of the star including the stellar disk fraction covered by heterogeneity features, the average temperature of the heterogeneous fraction, and the temperature of the stellar photosphere. Jointly with the three stellar properties, the methodology permits the retrieval of a host of exoplanet atmosphere properties including the chemical compositions and abundances, attributes of clouds and hazes, and the temperature profiles throughout the atmosphere. Our methodology is the first joint retrieval framework suited for the extraction of properties of exoplanetary atmospheres and their host stars. As such it sets a precedence for more detailed joint analysis techniques of exoplanets and their stars in the future.

We have applied our methodology to the transmission spectra of a sample of hot giant exoplanets to ascertain the influence of stellar heterogeneity and clouds/hazes in their spectra. The analysis distinguishes four groups of planets defined by the components needed to best explain their spectra. These four groups are illustrated in Table \ref{tab:condensed_expl}. In the first case, we find that the spectra of WASP-6b and WASP-39b are best fit with stellar heterogeneity as well as hazes and/or clouds. In the second case, there is marginal evidence for stellar heterogeneity effects and beyond substantial evidence for hazes and/or clouds in the spectra of HD 209458b and HAT-P-12b. The third group constitutes HAT-P-1b and WASP-31b and shows weak evidence against stellar heterogeneity but weak to substantial indications of clouds/hazes. In the fourth group three planets -- WASP-19b, WASP-17b, and WASP-12b -- are fit best by alkali and molecular absorbers with H$_2$ scattering without stellar heterogeneity and weak to no evidence of cloud/haze coverage. We emphasize that the suggestion against heterogeneity for WASP-19b is potentially due to the low data quality of the spectrum, and thus future studies of WASP-19b may suggest differently.

We note that joint retrievals of the stellar photosphere and exoplanetary atmosphere rely on the assumptions that the model components are reasonably correct and reasonably complete. Presently, however, even state-of-the-art models for stellar heterogeneity are based on very limited knowledge; furthermore, the hypothesized haze particles can provide similar spectral signatures in the optical. Thus, future efforts must collect better data to break the degeneracy between stellar contamination and possible atmospheric aerosols.

Ultimately, upcoming observatories will provide improved spectral resolutions and precisions useful for more definite and detailed joint analyses of transmission spectra. The simultaneous information on stellar and planetary properties facilitated through {\sc Aura} serves as a helpful tool in the analysis of present high-precision spectra and future high fidelity observations from the JWST and powerful ground-based facilities.

\section*{Acknowledgements}

AP is grateful for research funding from the Gates Cambridge Trust. AP thanks Siddharth Gandhi for contributing components of the retrieval code used in this work. AP acknowledges Ryan MacDonald for contributing aspects of the plotting routines. BR acknowledges support from the National Science Foundation Graduate Research Fellowship Program under Grant No. DGE-1143953. NM acknowledges support from the Science and Technology Facilities Council (STFC), UK. We thank Joanna Barstow and an anonymous reviewer for providing valuable comments on the manuscript. The results reported herein benefited from collaborations and/or information exchange within NASA'
s Nexus for Exoplanet System Science (NExSS) research coordination network sponsored by NASA's Science Mission Directorate. We thank the contributors of the Python Software Foundation and NASA's Astrophysics Data System.



\bibliographystyle{mnras}
\bibliography{references} 



\appendix \label{appendix_section}

\onecolumn
\section{Stellar Heterogeneity Parameters}\label{appendix_1}

\begin{table*}
\caption{Retrieved stellar heterogeneity parameters of the HMch models.}
\label{tab:stelparams}
\begin{tabular}{K{1.7cm} K{1.7cm} K{1.7cm} K{1.7cm}}
\hline
\hline
\rule{0pt}{3ex}  
Planet & $\delta$ &  $T_{\mr{het}}$ (K) & $T_{\mr{phot}}$ (K)\\ [1ex]
\hline
\rule{0pt}{3ex}
\\
 
 WASP-19b  & $0.06^{+0.07}_{-0.04}$  &  $5115^{+267}_{-472}$  & $5502^{+84}_{-81}$ \\[1ex]
 
 \\
 
 WASP-6b & $0.10^{+0.14}_{-0.03}$ & $4767^{+620}_{-314}$ & $5394^{+205}_{-61}$ \\[1ex]
 
 \\
 
 HD 209458b & $0.03^{+0.02}_{-0.01}$ & $3663^{+1998}_{-411}$ & $6088^{+89}_{-93}$ \\[1ex]
 
 \\
 
 HAT-P-1b  & $0.06^{+0.02}_{-0.02}$ & $3992^{+841}_{-566}$ & $5980^{+7}_{-7}$ \\[1ex]
 
 \\
 
 WASP-39b & $0.10^{+0.06}_{-0.04}$ & $4935^{+240}_{-309}$ & $5415^{+125}_{-125}$ \\[1ex]
 
 \\
 
 HAT-P-12b & $0.18^{+0.03}_{-0.02}$ & $4167^{+119}_{-157}$ & $4669^{+55}_{-48}$ \\[1ex]
 
 \\
 
 WASP-31b & $0.10^{+0.05}_{-0.04}$ & $5030^{+681}_{-1343}$ & $6297^{+83}_{-82}$ \\[1ex]
 
 \\
 
 WASP-12b & $0.04^{+0.02}_{-0.02}$ & $3986^{+1648}_{-692}$ & $6081^{+160}_{-157}$\\[1ex]
 
 \\
 
 WASP-17b & $0.06^{+0.05}_{-0.03}$ & $5848^{+784}_{-1232}$ & $6656^{+64}_{-69}$ \\[1ex]
 
 \\

\hline
\end{tabular}
\end{table*}


\bsp	
\label{lastpage}
\end{document}